\documentclass[aps,prx,twocolumn,superscriptaddress,noeprint,longbibliography, nofootinbib,floatfix]{revtex4}

\usepackage{graphicx}
\usepackage{bm}
\usepackage{amsmath, amssymb}
\usepackage{booktabs}
\usepackage{mathrsfs}
\usepackage[normalem]{ulem}
\usepackage{pifont}
\usepackage[mathscr]{euscript}

\usepackage{physics}
\usepackage{xcolor}

\newcommand{\micro}{$\mu{\rm m}$}

\begin{document}

\title{Anomalous magnetic noise in imperfect flat bands in the topological magnet  Dy$_2$Ti$_2$O$_7$ }

\author{Anjana M. Samarakoon}
\altaffiliation{Current address: Materials Science Division, Argonne National Laboratory, Argonne, IL, USA}
\address{Neutron Scattering Division, Oak Ridge National Laboratory, Oak Ridge, TN 37831, USA}
\address{Shull Wollan Center - A Joint Institute for Neutron Sciences, Oak Ridge National Laboratory, TN 37831. USA}	

\author{S. A. Grigera}
\email{sgrigera@fisica.unlp.edu.ar}
\address{Instituto de F\'{\i}sica de L\'{\i}quidos y Sistemas Biol\'ogicos, UNLP-CONICET, La Plata, Argentina}

\author{D. Alan Tennant}
\address{Neutron Scattering Division, Oak Ridge National Laboratory, Oak Ridge, TN 37831, USA}
\address{Quantum Science Center, Oak Ridge National Laboratory, Oak Ridge, TN 37831, USA}
\address{Shull Wollan Center - A Joint Institute for Neutron Sciences, Oak Ridge National Laboratory, TN 37831. USA}

\author{Alexander Kirste} 
\address{Physikalisch-Technische Bundesanstalt (PTB), 10587 Berlin, Germany}

\author{Bastian Klemke} 
\address{Helmholtz-Zentrum Berlin f{\"u}r Materialien und Energie, D-14109 Berlin, Germany.}

\author{Peter Strehlow}
\address{Physikalisch-Technische Bundesanstalt (PTB), 10587 Berlin, Germany}

\author{Michael Meissner}
\address{Helmholtz-Zentrum Berlin f{\"u}r Materialien und Energie, D-14109 Berlin, Germany.}

\author{Jonathan N. Hall\'en}
\address{TCM Group, Cavendish Laboratory, University of Cambridge, Cambridge CB3 0HE, UK}
\address{Max Planck Institute for the Physics of Complex Systems, 01187 Dresden, Germany}

\author{Ludovic Jaubert}
\address{CNRS, Universit\'e de Bordeaux, LOMA, UMR 5798, 33400 Talence, France}

\author{Claudio Castelnovo}
\email{cc726@cam.ac.uk }
\address{TCM Group, Cavendish Laboratory, University of Cambridge, Cambridge CB3 0HE, UK}

\author{Roderich Moessner}
\address{Max Planck Institute for the Physics of Complex Systems, 01187 Dresden, Germany}

\date{\today}

\begin{abstract}
The spin ice compound Dy$_2$Ti$_2$O$_7$ stands out as the first topological magnet in three dimensions, with its tell-tale emergent fractionalized magnetic monopole excitations. Its real-time dynamical properties have been an enigma from the very beginning. Using ultrasensitive, non-invasive SQUID measurements, we show that Dy$_2$Ti$_2$O$_7$ exhibits a highly anomalous noise spectrum, in three qualitatively different regimes: equilibrium spin ice, a `frozen' regime extending to ultra-low temperatures, as well as a high-temperature `anomalous' paramagnet. We show that in the simplest model of spin ice, the dynamics is {\it not} anomalous, and we present several distinct mechanisms which give rise to a coloured noise spectrum. 
In addition, we identify the structure of the {\it single-ion} dynamics as a crucial ingredient for any modelling.  
Thus, the dynamics of spin ice Dy$_2$Ti$_2$O$_7$ reflects the interplay of local dynamics with 
emergent topological degrees of freedom and a frustration-generated {\it imperfectly} flat energy landscape, and as such should be relevant for a broad class of magnetic materials.
\end{abstract}

\maketitle


\section{Introduction}

Spin ice materials are a paradigmatic example of three-dimensional topological behaviour~\cite{moessner_moore_2021}. Their prominence as a model system is supported by a remarkable level of quantitative agreement between experiment and relatively simple theoretical models, which has allowed for detailed understanding of the mechanisms underpinning their exotic equilibrium behaviour, in a way that is rare in strongly interacting many-body physics.

\begin{figure*}
\centering
\includegraphics[width=2.0\columnwidth]{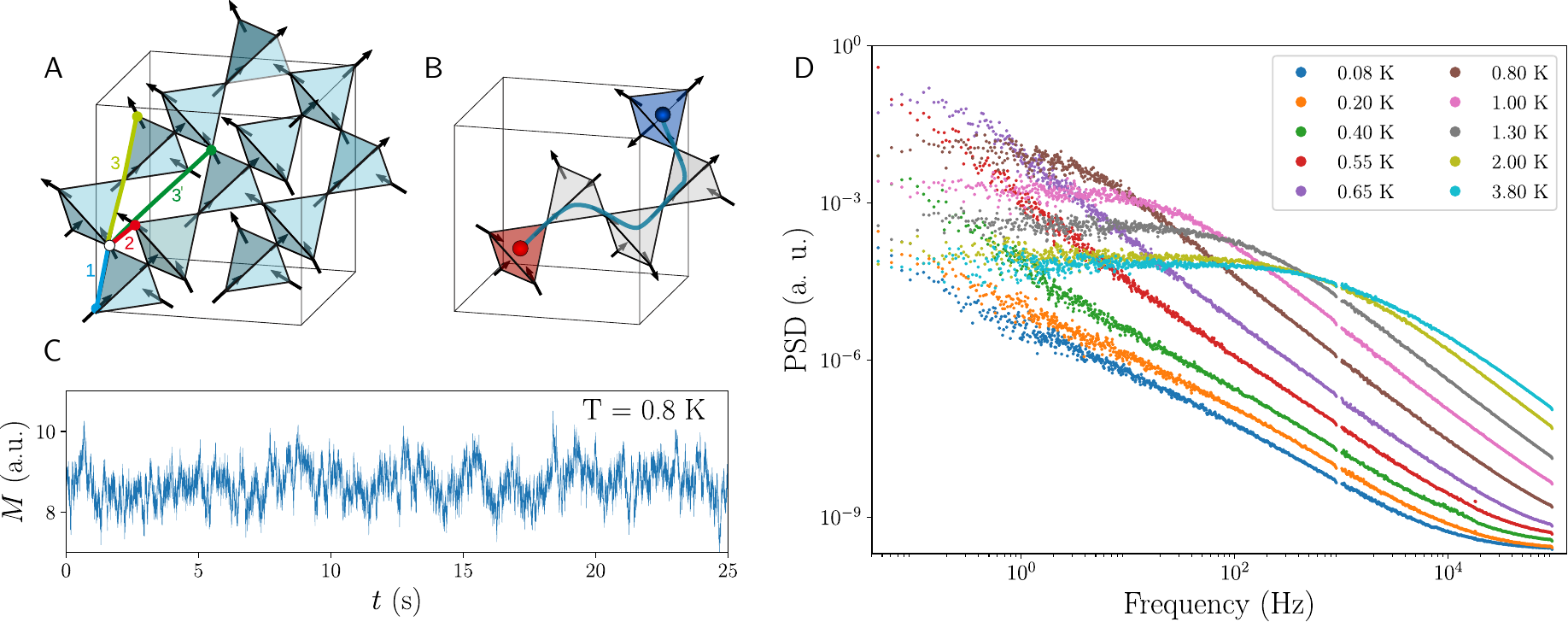}
\caption{Magnetic structure, monopole hopping and anomalous noise measurements in DTO. (A) The spins in DTO are located on a pyrochlore lattice. The magnetic interaction pathways are shown in color. These, together with long-range dipolar interactions, constrain the spin configurations to follow two-in two-out ice rules. (B) Breaking the ice rules results in the creation of a monopole anti-monopole pair which can separate leaving behind a Dirac string. Monopoles are constrained by the other spins to travel over a restricted manifold shown in grey. (C) As the monopoles move, they flip spins, detected as magnetic noise in the time domain. (D) The PSD signal for a selection of temperatures covering the full temperature range. The displayed temperatures are given in the legend. Each curve is composed of two data sets with different sampling rates (and hence different frequency windows), causing the gap in the PSD around $10^3$~Hz. 
}
\label{fig:trace}
\end{figure*}

Dy$_2$Ti$_2$O$_7$ (DTO) typifies topological spin ice behavior. It has a nearly flat band of low-lying spin configurations that satisfy ice rules, with magnetic monopole defects as excitations, Fig.~\ref{fig:trace}. The dynamics and annealing within the band are determined by the motion of these monopoles. When the band is highly entropic, spin ice has an elegant analogy with emergent electrostatics, and DTO shows this Coulomb phase over a range of temperatures.

Given the importance of spin models to the understanding of cooperative phases and transitions, it is all the more remarkable how enigmatic the dynamics of spin ice systems remains to date. Different experiments in  DTO~\cite{snyder2001spin,matsuhira2011spin,revell2013evidence,pomaranski2013absence} indicate a rapid slowdown of the dynamics upon lowering temperature. On a qualitative level, this can be understood by combining the well-established classical model Hamiltonian with a single-spin-flip dynamics appropriate for Ising systems~\cite{Ryzhkin,Jaubert2009,Jaubert2011}. Under the assumption of a unique and temperature-independent spin-flip time scale, e.g., due to single-ion quantum spin tunneling~\cite{tomasello2015single} (which we denote as $\tau_u$), a Lorentzian form for the magnetic susceptibility is predicted~\cite{Ryzhkin}. To leading order, this has a characteristic magnetic relaxation time scale obeying an Arrhenius law at low temperatures, inversely proportional to the monopole density, which is indeed activated~\cite{Jaubert2009}. This theory, however, fails to account for the non-Lorentzian shape of the curve, while quantitatively strongly underestimating the actual growth of the timescale (see Fig.~\ref{fig:Cole_fits} for a review of data in the literature), as well as failing to capture the nature of the irreversibility appearing around $T_\mathrm{irr}\approx0.6$ K. 

Here we address the origin of the anomalous cooperative dynamics, and whether it is intrinsic to spin ice. Attempts to explain such discrepancies resulted in models invoking finite-size effects, open boundary conditions, a temperature dependent time scale, and chemical substitution disorder~\cite{revell2013evidence,yaraskavitch2012spin,sala2014vacancy,Paulsen2014,Paulsen2016,paulsen2019nuclear}.
Recently, an intriguing analogy between spin ice phenomenology and generation/recombination noise in semiconductors was drawn based on high-temperature SQUID measurements~\cite{dusad2019magnetic}. In general, the anomalous behaviour of spin ice materials is often related to the physics of supercooled liquids and a possible avoided phase transition~\cite{matsuhira2001,matsuhira2011spin,matsuhira2011,kassner2015supercooled}. However, it is fair to say that a satisfactory understanding is still very much missing.

We report non-invasive SQUID noise measurements with unprecedented sensitivity and access to low temperatures (Fig.~\ref{fig:trace}), enlarging the experimental data for the dynamics across a broad range of temperatures and frequencies. We also present a detailed set of Monte Carlo simulations investigating different model families for spin ice systems. We show intrinsic anomalous dynamics in spin ice originating from memory effects of monopole motion in the imperfect flat band of Dy$_2$Ti$_2$O$_7$; but we also observe important contributions from more complex single-ion tunneling.


\section{SQUID noise measurements}

We use an ultrasensitive SQUID microsusceptometer to measure in a direct way magnetic noise across a broad temperature window from $0.08$~K to $4$~K. The experiment was set up in an adiabatic demagnetization refrigerator at the Physikalisch-Technische Bundesanstalt (PTB), Berlin. The SQUID-based setup allows measurement sensitivities of $10^{-15}$~tesla  in the magnetic field due to the sample (for details see Supplementary Note I).

A high-quality single crystal of Dy$_2$Ti$_2$O$_7$ was mounted on top of the SQUID sensor and the magnetic flux was recorded as a function of time, spanning the frequency range $\nu$ from $0.01$~Hz to $10^5$~Hz. The measurements were started at base temperature, and undertaken over a period of $24$~hours while the sample slowly warmed.

The single crystal was isotopically enriched to result in zero nuclear moment to ensure that the magnetic signal emanates exclusively from the electronic moments.


\section{Results}

Fig.~\ref{fig:trace}D summarises the results, displaying the noise power spectral density (PSD) from $0.08$~K to $3.8$~K. Two datasets were combined to improve the statistics. The experimental noise measurements cover about seven orders of magnitude in frequency and nine orders of magnitude in noise power, demonstrating the sensitivity and dynamical range of the apparatus. 

\begin{figure*}
\centering
\includegraphics[width=2\columnwidth]{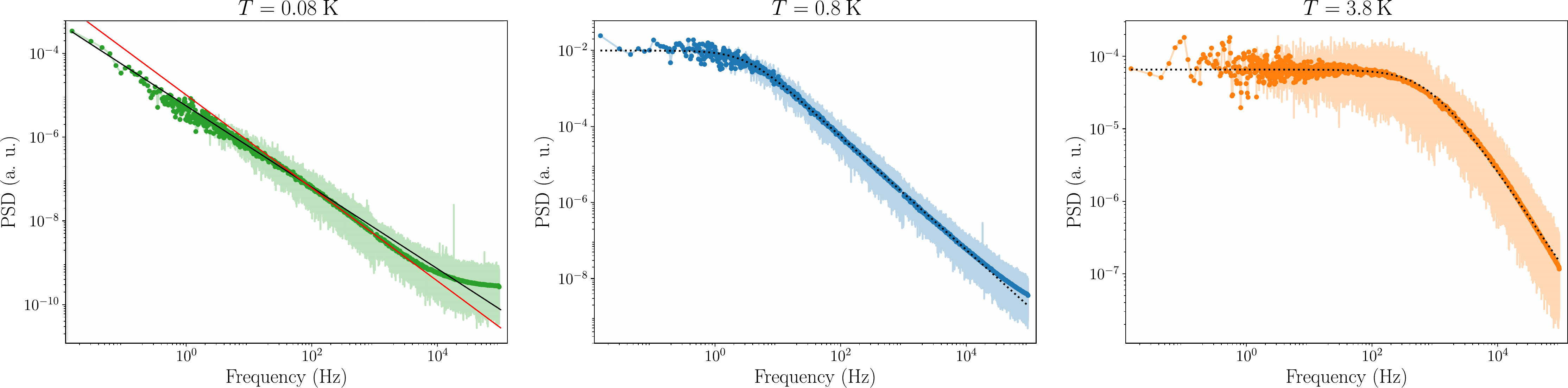} 
\caption{\label{fig:lowTT}\label{fig:intermediateTT}\label{fig:highTT}%
The raw (transparent lines) and window averaged (opaque points) PSD signal at three temperatures.
Left: At low temperatures, the PSD acquires an `S' shape suggestive of (at least) two distinct contributions at low and high frequency.
The black and red lines are guides to the eye and have approximate slopes $0.98$ and $1.12$ respectively. 
Middle: In the range between $750$~mK and $1.5$~K, a Cole-Cole form (dotted black line) fits the data well. 
Right: At higher temperatures ($\gtrsim 1.5$~K), the knee between plateau and scaling behaviour broadens. The high-frequency scaling regime is less clearly established and fitting for the exponent becomes more uncertain. 
The fitted values of the exponent $\alpha$ and the time scale $\tau$ are shown in Fig.~\ref{fig:Cole_fits}.}\
\end{figure*}\

At high temperature, the curves exhibit a plateau at low frequency and a decay at high frequency. These features are separated by a knee which moves to lower frequencies as the temperature is lowered until it is eventually squeezed out of the measurement window, while the total noise power also decreases. We distinguish three different temperature regimes. The high temperature {\em paramagnetic regime} crosses over to the {\em spin-ice regime} around $T \approx 1~{\rm K}$ as the spin-ice correlations gradually develop. As temperature is lowered below $T_{\rm irr} \approx  0.6 ~{\rm K}$, the irreversibility temperature of magnetisation measurements, the system enters a {\em non-equilibrium regime} characterised by extremely long magnetic relaxation times.
 
We start by focusing on the spin ice regime, which is of central interest to this study. The central panel of Fig.~\ref{fig:intermediateTT} shows the frequency dependence of the noise power at $T=0.8~{\rm K}$ alongside a fit to a Cole-Cole form 
\begin{equation}
    S_{\rm CC}(\nu)=\frac{A}{1+(2\pi\nu \tau)^\alpha}
    \, .
\label{eq:cole-colePSD}
\end{equation}
The fit works very well over the full frequency range; the main deviation is at the lowest frequencies, where the experimental data exhibit a weak rise rather than a perfect plateau. Two features stand out in the fit. Firstly, it covers a region of more than six orders of magnitude along both axes, i.e., frequency and noise power. Secondly, it exhibits an anomalous exponent, $\alpha\approx1.5$, as opposed to the $\alpha=2$ of a simple Lorentzian. More details on the data analysis and fitting procedure are given in the Sup. Info. 

Indeed, this is our first central experimental result: spin ice is well-known to be fully equilibrated at these temperatures, as hysteresis, history dependence and other signatures of out of equilibrium behaviour only set in around $T_{\rm irr}$. Nonetheless its noise spectrum exhibits a form otherwise familiar from the study of glasses and supercooled liquids. 

Turning towards the paramagnetic regime at higher temperatures, right panel of Fig.~\ref{fig:highTT}, the shape of the curve evolves slowly, with the knee softening somewhat. The dynamical range accessible in noise power decreases with the strength of the overall signal, and the uncertainty in the fitting parameters grows (see discussion in the Sup. Info.). 

Thirdly, in the non-equilibrium low temperature regime, the curve becomes more complex entirely. Furthermore, the absence of a plateau makes the fit to Eq.~\eqref{eq:cole-colePSD} somewhat poorly constrained. However, even in the accessible data window, it is apparent that there are (at least) two different portions with different anomalous slopes at intermediate and high frequency, with a possible further upturn (which however may be caused by the sensor noise), endowing the curve with an `S'-shape, see left-hand panel of Fig.~\ref{fig:lowTT}. 
A simple fitting form like the one used in equilibrium therefore no longer suffices. Additionally, the system is out of equilibrium and the details of the curve can depend on the preparation history of the measurement.

\begin{figure}
\centering
\includegraphics[width=0.9\columnwidth]{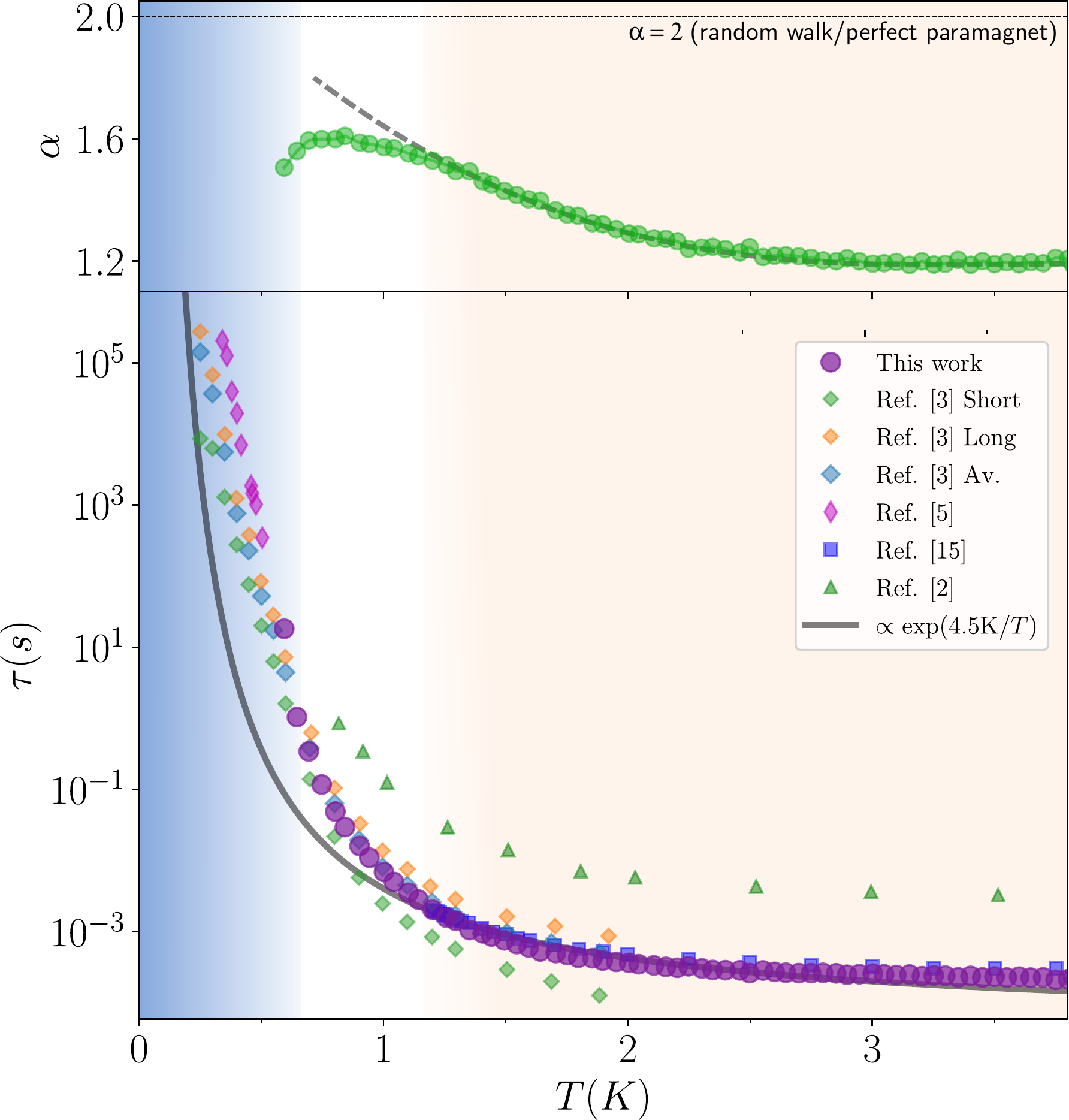}
\caption{ Anomalous exponent (top panel) and characteristic relaxation time scale $\tau$ (bottom) extracted from Cole-Cole fits to the experimentally measured PSD. The characteristic time is compared with values found in the literature using various techniques, and with an Arrhenius law. The shaded background indicates the three different temperature regimes: {\em paramagnetic} (orange), {\em spin ice} (white) and {\em non-equilibrium} (blue). The dashed line in the upper panel is an extrapolation of the high temperature behaviour of $\alpha$ into the spin ice regime. } 
\label{fig:Cole_fits}
\end{figure}

If one takes the parameters obtained from the equilibrium regimes at face value, two things are notable. First, the temperature dependence of the characteristic timescales $\tau(T)$ (purple circles in the lower panel of Fig.~\ref{fig:Cole_fits}) is close to those obtained previously by means of AC susceptibility measurements \cite{matsuhira2001}, specific heat \cite{pomaranski2013absence}, and high temperature SQUID noise experiments \cite{dusad2019magnetic} on single crystals. Second, the power $\alpha$ (upper panel in Fig.~\ref{fig:Cole_fits}) does not get close to $2$ (Lorentzian behaviour) at {\it any} temperature. Indeed, it seems to become even smaller, i.e., more anomalous, with increasing temperature.


\section{Modeling}

The highly non-Lorentzian relaxation is perhaps the most striking aspect of the dynamics in the spin ice regime where we know the system equilibrates and does not exhibit any glassy behaviour. The fact that it becomes less Lorentzian as the temperature is increased, where spin ice increasingly resembles a conventional paramagnet, is a separate surprise.

In order to determine the possible sources of this behaviour, we have conducted extensive simulations of the dynamics of spin ice. To do so, we have adopted the $\tau_u$-dynamics introduced above~\cite{Ryzhkin,Jaubert2009,Jaubert2011}. This is a stochastic model of incoherent dynamics, which only has the single timescale $\tau_u$ as input parameter. The timescale can be thought of as an effective (inverse) spin flip attempt rate that an isolated moment in spin ice would have. The actually observed spin flip rate then also takes into account the exchange field due to its interaction partners, via an acceptance probability $p=\min\{1,\exp(-\beta\, \Delta E)\}$ given by standard Metropolis dynamics at temperature $T$ for an energy difference $\Delta E$ between initial and final states. An appraisal of this assumption  {\it a posteriori} will form an important part of our discussion. 

We next present our analysis of such dynamics in a physically motivated set of spin ice models. These are based on the current best effective DTO Hamiltonian ${\cal H}_{\mathrm{OP}}$ ~\cite{Samarakoon_2020}. ${\cal H}_{\mathrm{OP}}$ involves strong nearest neighbor (nn) and dipolar interactions with weaker second and third neighbor interactions and reproduces equilibrium {as well as irreversible behavior}. Our further models are the simple nn spin ice model, ${\cal H}_{\mathrm{nn}}$; extension to include dipolar coupling, ${\cal H}_{\mathrm{dip}}$; and further addition of the third-neighbor interaction, ${\cal H}_{\mathrm{J_3^\prime}}$ (see Sup. Info. for detailed definitions). We finally contrast these to an unrestricted random-walk process for the monopoles yielding a straightforward Lorentzian behaviour. 
(We note that ${\cal H}_{\mathrm{dip}}$ and ${\cal H}_{\mathrm{OP}}$ undergo thermodynamic ordering transitions at  $T\approx 0.18 $~K.)

\begin{figure*}
\centering
\includegraphics[width=1.8\columnwidth]{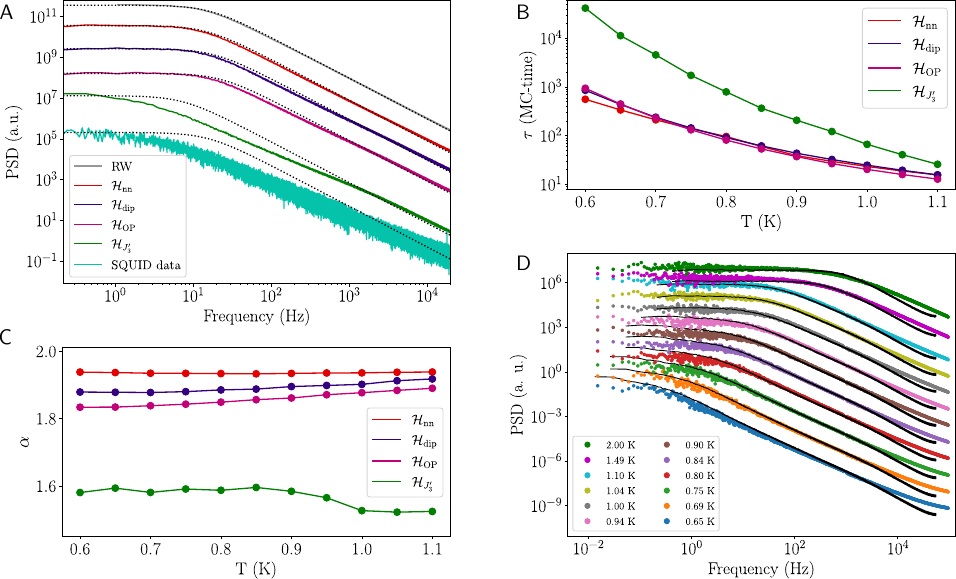}
\caption{{A)} PSD of (from top to bottom): a random walk on a diamond lattice, Monte Carlo simulations of the four spin ice Hamiltonians ${\cal H}_{\mathrm{nn}}$, ${\cal H}_{\mathrm{dip}}$, ${\cal H}_{\mathrm{OP}}$ and ${\cal H}_{J_3'}$ and experimental SQUID data at $T=0.8$K (shifted vertically for clarity). The spin ice Monte Carlo simulations were performed on a system of size $L=10$ with periodic boundary conditions and a $16$ spin cubic unit cell. The dotted black curves are fits of the form $A/(1+\tau^2 \nu^2)$, showing how the models display more anomalous decay as we move down the plot.
{ B)} Characteristic relaxation time scale $\tau$, and {C)} anomalous exponent for the model Hamiltonians ${\cal H}_{\mathrm{nn}}$, ${\cal H}_{\mathrm{dip}}$, ${\cal H}_{\mathrm{OP}}$ and ${\cal H}_{\mathrm{J_3^\prime}}$. Parameters extracted from Cole-Cole fits to Monte Carlo data (see Sup. Info. for details). {D)} Comparison of the experimental curves (symbols) and Monte Carlo simulations (black lines) for the Hamiltonian ${\cal H}_{\mathrm{J_3^\prime}}$. Temperatures from top to bottom are listed in the legend; the PSD curves are shifted vertically for clarity.}
\label{fig:modelcomparison}
\end{figure*}

We first compare the behaviour of these models with experiment in the equilibrated spin ice regime ($T=0.8$ K),  Fig.~\ref{fig:modelcomparison}. It is immediately apparent that the random walk produces essentially perfect Lorentzian behaviour,  with ${\cal H}_{\mathrm{nn}}$ likewise only deviating by an almost imperceptible amount matching theoretical expectations. Surprisingly, anomalies become apparent in ${\cal H}_{\mathrm{dip}}$ and even more so for ${\cal H}_{\mathrm{OP}}$. $\alpha$ is in fact quite tunable, and it drifts considerably as the strength of the further-neighbour interaction ${J_3^\prime}$ is varied (see Sup. Info.). For ${\cal H}_{J_3^\prime}$ the best match to the data ($J_3^\prime=0.4$~K) results in strong anomalous behaviour (see panels C and D of Fig.~\ref{fig:modelcomparison}).

Crucially, we have thus identified a group of quite simple, disorder-free model Hamiltonians which exhibit variably anomalous noise, implying a broad distribution of timescales, even though the microscopic spin dynamics is parameterized by a single rate.

We next address the high-temperature regime, only briefly as there is a fine recent pioneering study devoted to this, which studied the noise in spin ice down to a temperature $T=1.2$~K~\cite{dusad2019magnetic}. 
In this regime, the narrow range in frequency can artificially lower the value of $\alpha$ (see Supplementary notes IV and VI). 
Be the uncertainty in the fitting procedure as it may, the central finding is that experiments depart significantly from a Lorentzian behaviour at high temperatures, whereas Monte Carlo simulations do not. 

The root of this departure must therefore lie in features not included in the model with $\tau_u$ dynamics. As argued below, it would seem rather natural that more complex spin flip behaviour,  rather than a purely cooperative effect of the spins, plays a role here, with an interplay of phonons, higher crystal field levels and a broadening distribution of local environments standing out as likely culprits. 

We finally turn to the low-temperature regime, where spin ice falls out of equilibrium at $T_{\rm irr}\approx 0.6~{\rm K}$. This regime has been extensively studied, using neutron scattering and susceptibility measurements~\cite{snyder2001spin,snyder2004low,matsuhira2011spin,quilliam2011dynamics,yaraskavitch2012spin} as well as a number of time-dependent protocols for, e.g., the magnetisation, aimed at eliciting details of the non-equilibrium behaviour~\cite{orendavc2007magnetocaloric,slobinsky2010unconventional,Paulsen2014,kaiser2015ac,giblin2011creation}. Although the dynamical origin of this freezing is still not very well understood, the irreversibility is captured by ${\cal H}_{\mathrm{OP}}$ \cite{Samarakoon_2020}. 
Thanks to the wide temperature and frequency range our data provide some new insights: Firstly, as expected, the knee terminating the low-frequency plateau moves towards low frequencies upon cooling. This simply reflects the well-known slowing down of the dynamics in spin ice~\cite{snyder2001spin} as the monopoles become sparser~\cite{Ryzhkin}. The latter is also evident in the noticeable decrease of the total noise power as the temperature is lowered below $T_{\rm irr}$.

Additional structure emerges in the noise spectrum as can occur when there are two (parametrically) distinct timescales present. Most simply two superposed Lorentzians, with the slow one corresponding to a larger signal, would yield an S-shape profile as observed. It is tempting to speculate that here we have on the one hand the relatively fast motion of monopoles trapped in their individual surroundings, where however the confined paths cannot lead to a large change of the magnetisation; and on the other hand rare longer-distance excursions of monopoles, towards e.g., another local trap.  Naturally, one can extend such an argument to a phenomenology based on an ensemble of Lorentzians with a distribution of characteristic timescales chosen to fit experiment~\cite{Ryzhkin}.


\section{Discussion}

Taken together, the theoretical modelling situation can be perhaps succinctly summarised as follows. An entirely memory-free, unconstrained random motion of monopoles  with a single, temperature independent flipping rate, $\tau_u$, yields the normal (Lorentzian) power $\alpha=2$. This, however, is modified by two mechanisms originating in single-ion and cooperative effects, with opposite predominance at high and low temperatures. 

The cooperative effects have several origins. First, the Dirac strings lead to anticorrelations in the monopole hops; their effect on $\alpha$, however, turns out to be tiny. 
Next are the long-range dipolar interactions, which can be meaningfully broken down ~\cite{Isakov2005} into its (`projectively equivalent') Coulomb component, which preserves a flat energy landscape for the ground states, and quadrupolar and higher order corrections, which do not. We find that the contribution due to the former is also tiny (not shown), whereas the latter leads to a visibly anomalous behaviour, albeit still far from that observed experimentally. ${\cal H}_{\mathrm{OP}}$ and ${\cal H}_{\mathrm{J_3^\prime}}$ then include additional features in the form of further-neighbour interactions, yielding yet more anomalous behaviour. 

Therefore, frustration yields an {\it approximately flat energy landscape}, in which the spin ice regime with its monopoles as topological defects arises. Due to this flatness, the monopoles remain mobile and their motion produces a noise signal over a broad temperature range. Thanks to our sensitive low-$T$ measurements, this signal remains detectable even in the regime where the monopoles become sparse. 

In a featureless energy landscape, the monopole motion is well approximated by a featureless 'Lorentzian' walk, $\alpha\approx2$.
It is when the energy landscape becomes complex in the presence of perturbations away from the ideal spin ice model that the anomalous behaviour sets in. Interestingly, our results suggest that entropic and energetic contributions that retain the degeneracy between the ice states, and hence do not lead to ordering, are also least  (if at all) effective at producing anomalous behaviour. By contrast, the progressive increase in anomalous behaviour suggests that the leading effect on the noise is due to interactions that ultimately cause spin ice to order, and it is evident {\it even far above the corresponding ordering temperature} -- the `opposite' of supercooling, as it were. 

This combination of frustration-induced flatness and perturbation-induced complexity of the landscape we believe should be a general conceptual framework for understanding the anomalous noise in topological magnets.  

Returning to the actual material, neither ${\cal H}_{\mathrm{OP}}$ , nor ${\cal H}_{J_3^\prime}$ can accurately describe the thermodynamics and dynamical behavior of DTO.  ${\cal H}_{\mathrm{OP}}$ has its origin in a detailed machine-learning based analysis of equilibrium neutron scattering, susceptibility and specific heat data and can accurately reproduce a number of thermodynamic properties of DTO beyond those used for its training. It qualitatively shows the expected anomalous noise behaviour, but quantitatively falls short of the mark. Variations on its parameters to better fit the noise pattern result in ${\cal H}_{J_3^\prime}$ at the expense of the correct thermodynamical description of the material.  Furthermore, none  of  the  simulations  capture  the high-temperature experimental behaviour, that becomes progressively  more,  rather  than  less, anomalous as temperature  is  increased. This is surprising, from the perspective of the $\tau_u$ model, whose single spin flip dynamics (appropriate for a paramagnet) yields a simple Lorentzian at high temperatures.

Let us revisit the central dynamical assumption of a single, temperature independent flipping rate encoded by $\tau_u$. The underlying spin flip process involves the flipping of a large spin with a considerable Ising barrier, in the presence of a `bath' of phonons with temperature-dependent occupancies and structure in its density of states~\cite{tomasello2015single,Ruminy2017}. The Ising barrier itself microscopically derives from a complex crystal-field level scheme, which in turn allows various flipping paths, involving activation over, or tunnelling through, the barrier. Their respective rates will in general depend on temperature, allowing a complex temperature dependence of the resulting net rate. In addition, these flip rates depend on the local spin configurations, e.g., via the local distribution of transverse fields providing effective matrix elements between the crystal field levels. As the temperature rises, more flipping paths contribute, and the distribution of local environments broadens both spatially and temporally, so that  one would expect increasing complexity of the resulting dynamics.

Turning to the signatures of complex spin-flip dynamics in detail, we note that 
already at low temperature, there is a divergence of the dynamical timescale $\tau(T)$ extracted from experiment which is in excess of the cost of an isolated monopole, generally expected to set the single-spin flip time scale~\cite{Ryzhkin}. The simplest reason for such a discrepancy would be an autonomous Arrhenius law of the effective spin flip attempt rate, rather than a cooperative effect. 

In turn, the anomalous behaviour of  $\alpha$ in DTO (Fig. \ref{fig:Cole_fits})  can  be understood in terms of two counterveiling mechanisms: as the temperature is lowered starting in the paramagnetic regime, the anomaly with its origin in the spin-flip dynamics decreases, with the corresponding increase in $\alpha$, which should eventually reach $\alpha=2$ (dashed line in Fig.~\ref{fig:Cole_fits}) were it not counteracted by the second effect, the gradual onset of cooperative effects within the spin model. Together, these opposing tendencies  lead to a levelling off of $\alpha$, before in the frozen phase an analysis in terms of a single exponent no longer accounts for the complexity of the anomalous behaviour entirely.

In this picture, the size of the anomaly has two sizeable contributions at 0.8 K, a spin-flip and a cooperative one. Absent a detailed analysis of the former, their respective sizes are not available quantitatively. An initial starting assumption suggested by the data is for both contributions to be roughly of equal size in this regime. In that case, the size of the anomaly observed in simulations of ${\cal H}_{\mathrm{OP}}$ would actually be consistent with the experimental results.

This scenario has the attraction that it fits all the thermodynamic experimental data -- heat capacity, neutron scattering, noise, magnetic susceptibility, magnetisation, etc. -- within the purview of ${\cal H}_{\mathrm{OP}}$; but it still requires  future work to obtain a detailed description of the single ion dynamics needed to complete the numerical modelling of the dynamic properties of these systems.

Independently of this sharpening of our understanding of the modelling of the DTO spin ice material, a central theoretical insight is that the anomalous behaviour, encoded by the downturn of $\alpha$, can arise cooperatively -- but not entirely straightforwardly. In particular, our simulations demonstrate that it is a phenomenon due to corrections beyond the Coulomb phase description: neither Dirac strings nor Coulomb interactions between monopoles produce a sizeable anomaly, but it is rather farther range interactions which endow the energy landscape for monopole motion with additional structure that make a large contribution. The resulting behaviour resembles a supercooled liquid but it can evidently happen {\em above} any thermodynamic transition temperature. 

From a more conceptual perspective, a finite-frequency response, such as the one probed here, will inevitably be sensitive to a combination of universal behaviour -- such as phase ordering -- and non-universal microscopic details. Both turn out to be very interesting in spin ice materials.


\section{Conclusion}
Our ultrasensitive SQUID study reveals many facets of anomalous dynamics in Dy$_2$Ti$_2$O$_7$. Frustration yields an unusual topological magnetic state supporting magnetic monopole excitations. While the simplest nearest-neighbour and dipolar spin ice models show (close to) Lorentzian behaviour, experiments as well as more realistic model Hamiltonians show evidence of intrinsic anomalous dynamics. We identify a family of models that show how perturbations which generate a complex energy landscape result in memory effects. Although supercooled-like, this robustly non-Lorentzian behaviour can occur as a {\it precursor} 
far above the actual ordering. Further, compelling evidence for complex spin-flip dynamics contributing to the anomalous behaviour, most strikingly at high temperature, is also given. 


\section*{Acknowledgements}
This work was partly supported by the Deutsche Forschungsgemeinschaft under grants SFB 1143 (project-id 247310070) and the cluster of excellence ct.qmat (EXC 2147, project-id 390858490), by the Engineering and Physical Sciences Research Council (EPSRC) Grants No. EP/K028960/1, No. EP/P034616/1, and No. EP/T028580/1 (CC), the Agence Nationale de la Recherche, under Grant No. ANR-18-CE30-0011-01, and Agencia Nacional de Promoci\'on Cient\'\i fica y Tecnol\'ogica through PICT 2017-2347. Part of this work was carried out within the framework of a Max-Planck independent research group on strongly correlated systems. The computer modeling used resources of the Oak Ridge Leadership Computing Facility, which is supported by the Office of Science of the U.S. Department of Energy under contract no. DE-AC05-00OR22725. AMS and DAT acknowledge the support from the US DOE office of scientific user facilities. This material is based upon work supported by the U.S. Department of Energy, Office of Science, National Quantum Information Science Research Centers, Quantum Science Center.

\section*{Author contributions}

Conception of measurement (PS MM) and wider project (DAT SAG CC RM). Noise measurements (AK BK). Modelling, analysis and interpretation (AS SAG DAT JH LJ CC RM). Paper writing (RM CC SAG DAT with input from all coauthors).

\bibliography{references}

\newpage

\setcounter{section}{0}
\setcounter{figure}{0}

\renewcommand{\thesection}{S\Roman{section}}
\renewcommand{\thefigure}{S\arabic{figure}}

\section*{Supplementary Material}


\section{Technical details about the SQUID susceptometer setup}

\begin{figure}
\includegraphics[width=\columnwidth]{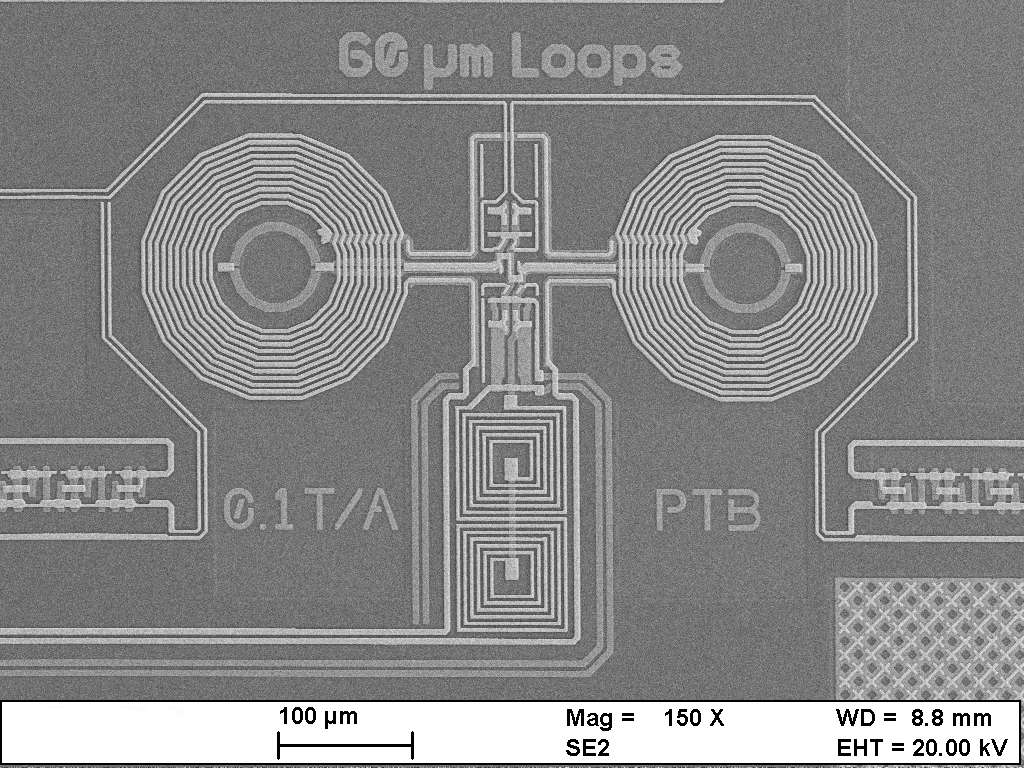}
\caption{SEM image of a microsusceptometer consisting of two $60~ \mu{\rm m}$  diameter loops connected in series as part of the SQUID circuit, each of them surrounded by an 11-turn field coil. The center-to-center separation between the loops (baseline of the first-order gradiometer) is  350 \micro .}
\label{S1}
\end{figure}

Sensitive measurements of magnetic noise have been performed using a SQUID microsusceptometer, a device that is designed with microscale dimensions and fabricated using reliable multilayer thin-film processes. It integrates gradiometric pick-up loops to detect the signal of a sample, field coils to produce an excitation magnetic field at the location of the sample, and a superconducting quantum interference device (SQUID) to read out the magnetic flux from the pick-up loops. We used a microsusceptometer of the type ``C6 SM'' designed and produced at PTB as part of the mask set ``C6'' in 2010 \cite{drung2014thin}. This device uses a first-order SQUID series gradiometer made up of two circular pick-up loops with a diameter of 60 \micro\ and a distance (baseline) of 350 \micro . Fig.~\ref{S1} shows an SEM image of the main parts of the microsusceptometer, which is hosted on a $3.3 \times 3.3~ {\rm mm}^2$ chip also carrying a SQUID current sensor. The susceptometer SQUID was read out by a second-stage SQUID array (PTB type ``C6 X216FB'') on a separate chip, followed by a room-temperature flux-locked loop (FLL) electronics (type Magnicon XXF-1). The main advantage of this two-stage SQUID readout is to minimize the effective intrinsic noise, as the noise contribution from the SQUID electronics is reduced.

\begin{figure}
\includegraphics[width=\columnwidth]{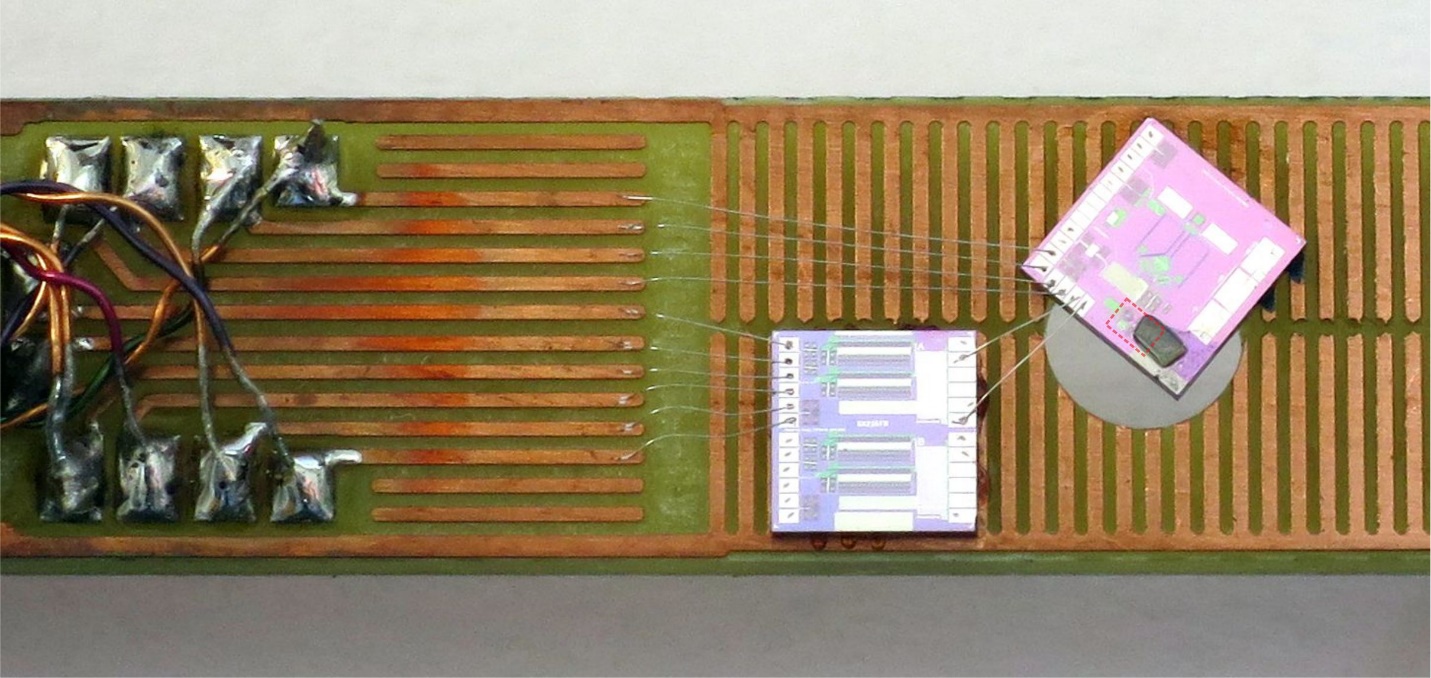}
\caption{Part of the susceptometer setup with two $3.3 \times 3.3~ {\rm mm}^2$ chips assembled on a printed circuit board (PCB). The actual microsusceptometer is centered over the hole in the PCB and partly covered by the sample, which is appearing in dark gray color. The area enclosed by the red rectangle corresponds to the image shown in Fig.~\ref{S1}.}
\label{S2}
\end{figure}

Both SQUID chips were integrated on a printed circuit board (PCB) (FR4, 1.6 mm thick). Its copper surface layers were structured into fine combs to maintain a good heat conduction, but to reduce eddy currents as well as thermal magnetic noise. For this reason, the actual susceptometer was placed above a hole in the PCB. All these components were enclosed in a superconducting shield made of Nb, its central part being a tube with an inner diameter $d_{\rm shield} = 9.2~ {\rm mm}$ and a length  $l_{\rm shield} = 68~ {\rm mm}$. The Dy$_2$Ti$_2$O$_7$ sample investigated was prepared from a single crystal in the form of a slab (0.2 mm thick) with approximate lateral dimensions of $0.9 \times 0.5 {\rm mm}^2$. It was mounted on the microsusceptometer with Apiezon N grease, covering completely pick-up loop and field coil at one side of the susceptometer, see Fig.~\ref{S2}.
For the noise measurements on a sample (noise as a signal) it is important not only to minimize intrinsic noise from the susceptometer, but also to suppress the coupling to external magnetic fields. The residual sensitivity is determined by (the product of) three quantities: (a) the imbalance of the first-order SQUID gradiometer (better than 0.35 \%, normalized to the area of a single pick-up loop, $A_{\rm p}$), (b) the shielding of the superconducting enclosing (depending on the aspect ratio of the tube, $d_{\rm shield}/l_{\rm shield}$), and (c) the shielding of a few layers of high-permeability magnetic shielding foil around the superconducting enclosing. In addition to that, the gradiometer balance with respect to the applied field produced by the integrated field coils (better than 0.11 \%, normalized to $A_{\rm p}$) counts to diminish background signals.

The SQUID susceptometer was operated in an adiabatic demagnetization refrigerator (ADR) capable of reaching temperatures as low as 75 mK. It was based on a dewar from Infrared Laboratories (similar to the HDL series) with liquid nitrogen and liquid helium pre-cooling, incorporated a bi-stable mechanical heat switch, a superconducting magnet surrounded by a magnetic high-permeability shield to reduce stray fields and a paramagnetic salt pill unit with two stages held in place by a string suspension. The first stage of the salt pill unit contained about 143 g of the gadolinium gallium garnet Gd$_3$Ga$_5$O$_{12}$ (GGG) and was used for heat sinking purposes at an intermediate temperature. The second stage contained about 50 g of ferric ammonium alum Fe$_2$(NH$_4$)$_2$(SO$_4$)$_4$24H$_2$O (FAA) and held the platform carrying the susceptometer setup. The superconducting solenoid was not equipped with a persistent switch so that the finite field for a given temperature required permanent current supply and control by the external power supply. To cover a wide range of temperatures with minimum fluctuations, we decided to cool down the ADR to the lowest temperature and measure during the natural warm-up caused by the intrinsic heat leak. The warming rate is then determined by the (variable) intrinsic heat leak to the FAA stage and the temperature dependent heat capacities of its main components (FAA, Cu). We can specify this warm-up by the relative warming rate $(dT/dt)/T$, for which we observed $1.5 \times 10^{-5}$~s$^{-1}$ at 80 mK, a maximum of $4.3 \times 10^{-5}$~s$^{-1}$ occurring at 570 mK, and $3.4 \times 10^{-5}$~s$^{-1}$ at 1 K followed by further dropping values up to 4 K. Another advantage of this practice with zero coil current (and disconnected external power supply) is that it avoids any potential influence from a varying stray field of the magnet.

\begin{figure}
    \centering
    \includegraphics[width=\columnwidth]{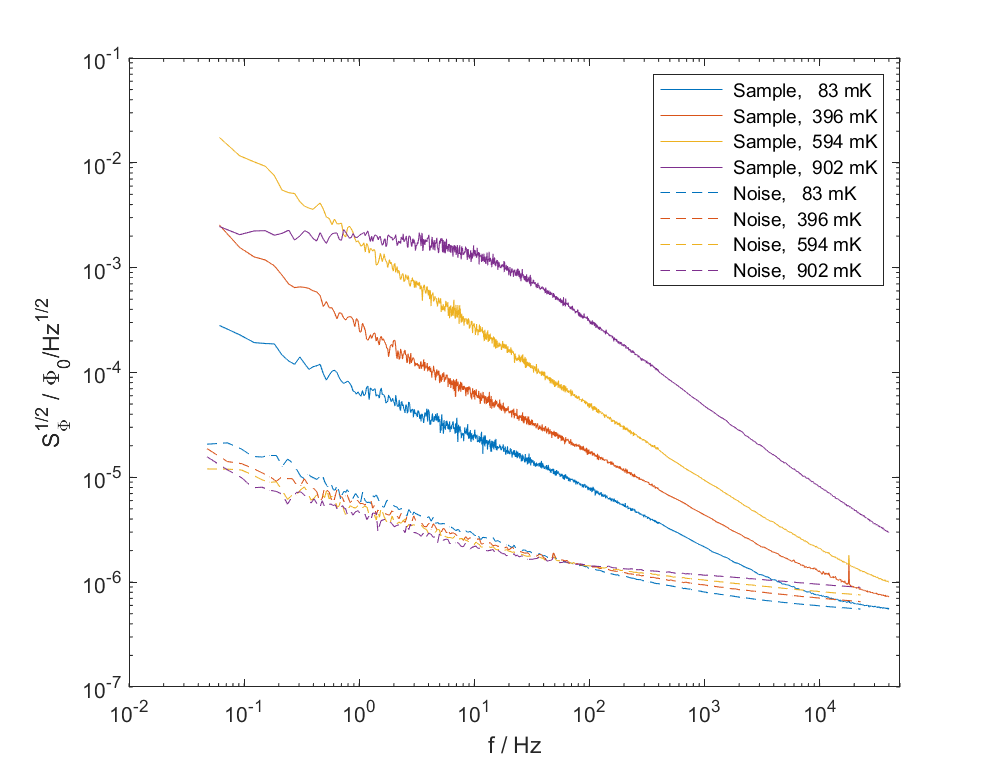}
    \caption{Flux noise ($S_\Phi^{1/2}$, corresponding to the square root of the PSD) measured with the SQUID susceptometer setup with Dy$_2$Ti$_2$O$_7$ sample (full lines) and without sample (dashed lines), respectively, at various temperatures.}
    \label{S3}
\end{figure}

The output voltage from the SQUID electronics was sampled using 24-bit data acquisition cards with integrated anti-aliasing filters (National Instruments PCI-4461 and NI PCI-4462) at sample rates up to 200 kSa/s.
The intrinsic noise of the susceptometer setup was measured in an independent run with the empty microsusceptometer, i.e., without any sample. Fig.~\ref{S3} shows this intrinsic flux noise $S_\Phi^{1/2}(f)$ at various temperatures in comparison with the flux noise obtained in the sample measurement. At 80 mK, the white noise is equal to $\approx 0.5 ~\mu \Phi_0/{\rm Hz}^{1/2}$, $(\Phi_0 = h/2e \approx 2.068 \times 10^{-15}$ Vs being the magnetic flux quantum), while the noise increases to $\approx 19~ \mu\Phi_0/{\rm Hz}^{1/2}$ at 0.1 Hz. For increasing temperature, we observe the usual increase of the white noise as well as a decrease of the 1/f noise component at low frequencies. As is visible in Fig. \ref{S3} for temperatures below 600 mK, the sample signal approaches the intrinsic white noise level at the highest frequencies (above $\sim 10$ kHz), which results in a slight upturn of $S_\Phi^{1/2}(f)$. However, at low frequencies ($<100 {\rm Hz}$), the intrinsic noise can be neglected at all temperatures investigated in this study.

\section{Model Hamiltonians}
\label{sec:modham}

The simplest model considered here is the standard nearest-neighbour spin ice Hamiltonian
\begin{equation}
    \label{eq:H_NN}
    {\cal H}_{\rm{nn}} = -J_{\rm{eff}}\sum_{\left<i,j\right>}\Vec{S}_i \cdot \Vec{S}_j \, ,
\end{equation}
where the sum is over nearest neighbour pairs. The nearest-neighbour model is an approximation of the dipolar Hamiltonian, which consists of long-range dipolar spin interactions and antiferromagnetic nearest-neighbour exchange, 
\begin{eqnarray}
\label{eq:H_dip}
{\cal H}_{\rm{dip}}\   \!\!\! &=& \!\!\! \  Da^3 \sum_{i<j}\left[
\frac{\Vec{S}_i \cdot \Vec{S}_j}{r_{ij}^3} - \frac{3 \left( \Vec{S}_i \cdot \Vec{r}_{ij}\right) \left( \Vec{S}_j \cdot \Vec{r}_{ij}\right)}{r_{ij}^5}  \right]
\nonumber \\ 
\!\!\! &+& \!\!\!  \  J_1\sum_{\left<i,j\right>}  \Vec{S}_i \cdot \Vec{S}_j 
\, , 
\end{eqnarray}
where $D$ is the dipolar coupling constant and $a$ is the spin nearest-neighbour distance. The exchange strength $J_{\rm{eff}}$ used in the simulations was chosen as a function of temperature so that the monopole densities of ${\cal H}_{\rm{nn}}$ and ${\cal H}_{\rm{dip}}$ match as well as possible across the temperature range of interest. This was done as our numerical analysis indicates that the monopole density is the dominant parameter controlling the behaviour of the noise observed in both ${\cal H}_{\rm{nn}}$ and ${\cal H}_{\rm{dip}}$.

\begin{figure}
    \centering
    \includegraphics[width=\columnwidth]{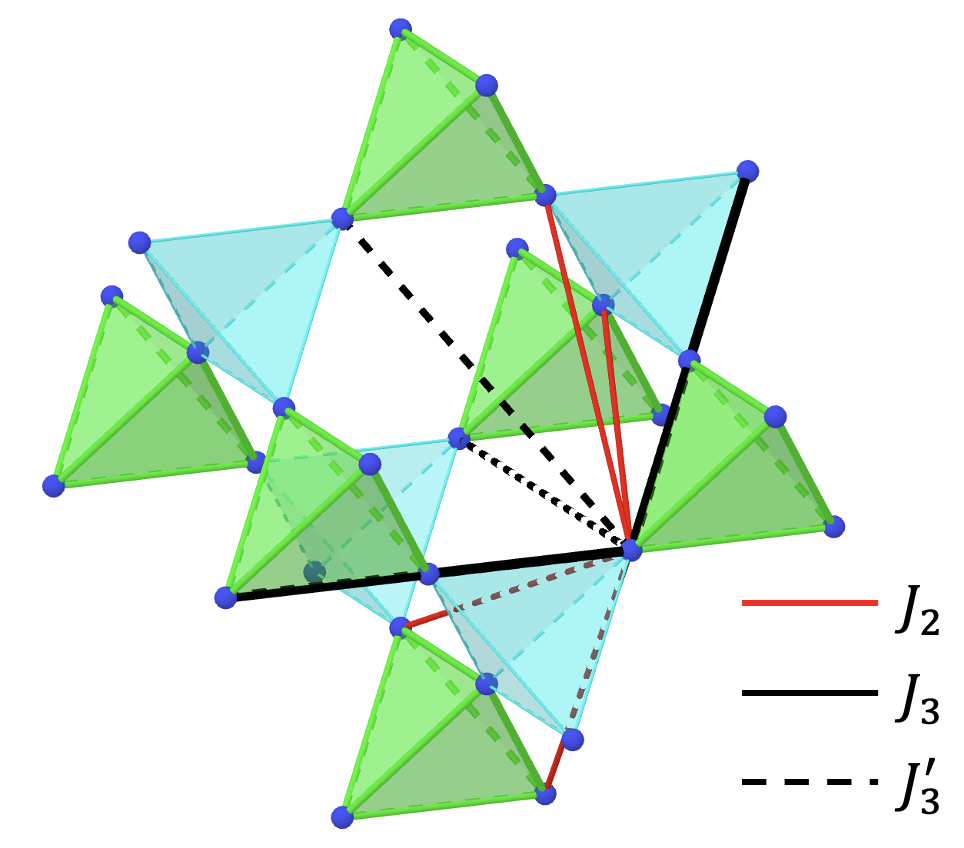}
    \caption{Section of the pyrochlore lattice with the two types of tetrahedra shown in green and light blue, and the spin positions marked as dark blue spheres. The red lines indicate second-neighbour exchange ($J_2$) for a specific spin, and the solid and dashed black lines indicate the two types of third-neighbour exchange ($J_3$ and $J_3'$, respectively).}
    \label{fig:J3+J3primeSketch}
\end{figure}

We further consider two additional Hamiltonians ${\cal H}_{\rm{OP}}$ and ${\cal H}_{J_3'}$, where further-neighbour exchange terms were included, connecting each spin to their second- and third-neighbours. The second-neighbour exchange has strength $J_2$ and connects each spin to twelve others. There are two types of third-neighbour interactions, each connecting a spin to six of its neighbours. The first of these has strength $J_3$ and lies parallel to the nearest-neighbour interactions. The second has strength $J_3'$ and connects spins across the hexagonal loops. These interactions are shown in Fig.~\ref{fig:J3+J3primeSketch}, and the resulting Hamiltonian is
\begin{eqnarray}
\label{eq:H_comp}
{\cal H}\   \!\!\! &=& \!\!\! \  Da^3 \sum_{i<j}\left[
\frac{\Vec{S}_i \cdot \Vec{S}_j}{r_{ij}^3} - \frac{3 \left( \Vec{S}_i \cdot \Vec{r}_{ij}\right) \left( \Vec{S}_j \cdot \Vec{r}_{ij}\right)}{r_{ij}^5}  \right]
\nonumber \\ 
\!\!\! &+& \!\!\!  \  J_1\sum_{\left<i,j\right>}  \Vec{S}_i \cdot \Vec{S}_j + J_2 \sum_{\left<i,j\right>_{2}}  \Vec{S}_i \cdot \Vec{S}_j 
\nonumber \\
\!\!\! &+& \!\!\! J_3\sum_{\left<i,j\right>_3}  \Vec{S}_i \cdot \Vec{S}_j  
+ J_3'\sum_{\left<i,j\right>_{3'}}  \Vec{S}_i \cdot \Vec{S}_j 
\, .
\nonumber \\
\!\!\! &\  &
\end{eqnarray}

In the three Hamiltonians with dipolar interactions, we set the interaction strength $D=1.3224 \ \rm{K}/ a^3$ and the nearest neighbour exchange $J_1=3.41$~K. ${\cal H}_{\rm{OP}}$ is optimised to reproduce neutron scattering, magnetic susceptibility and specific heat experimental data, and has further-neighbour exchange strengths $J_2=0.0$~K, $J_3=-0.00466$~K and $J_3'=0.0439$~K. ${\cal H}_{J_3'}$ is chosen to match the SQUID noise measurements as well as we could, and only has further-neighbour exchange across the hexagons, namely $J_2=J_3=0$~K and $J_3'=0.4$~K. 
%
%

\section{Numerical simulations}
\label{sec:numdet}

All results presented here were obtained from Monte Carlo simulations using the standard Metropolis algorithm. A system of linear size $L=10$ with periodic boundary conditions and a $16$ spin cubic unit cell was used, corresponding to $N_s = 16\times10^3$ spins. The dipolar interactions were included using the Ewald summation technique. $10^4$ to $10^5$ cooling steps ($1$ step $= N_s$ spin flip attempts) were used to equilibrate the system, depending on the measurement temperature. The net magnetisation of the system was then measured at fixed temperature, with a sampling rate of $16$ measurements per MC step for a total time window of between $3.3\times 10^4$ and $4.0\times10^6$ MC steps. Because the plateau extends to larger frequencies as the temperature increases, a shorter time window was sufficient at the higher temperatures. The PSD was calculated from the magnetisation data using Welch's method of power spectra estimation. 

To ensure that the cooling process was sufficient to reach equilibrium, a small number of control measurements were performed at the lowest temperature $T=600$~mK. The system was allowed to evolve at this temperature for an additional $10^6$ MC steps before the magnetisation was measured as described above. The PSD calculated from these simulations showed no significant difference from the one calculated after the standard $10^5$ cooling steps. 

The number of monopoles present in the system fluctuates naturally about an average value that is a function of temperature. To avoid significant finite size effects, the simulations were kept at temperatures where a non-vanishing density of monopoles was present at all times throughout the measurement. For a system of size $L=10$ as used here, this gives us a minimum computationally accessible temperature of approximately $600$~mK.  


\section{Fitting procedure}
\label{sec:methods}
\label{sec:numdet_fitting}

We performed both Cole-Cole and Davidson-Cole fits, 
\begin{equation}
    S_{\rm CC}(\nu)=\frac{A}{1+(2\pi\nu \tau)^\alpha}
\, , \quad 
    S_{\rm DC}(\nu)=\frac{A}{(1+2\pi\nu \tau)^\alpha}
    \, .
\label{appeq:cole-colePSD}
\end{equation}
using linear regression to the logged window averaged PSD. Three fitting parameters were used: plateau height, time scale and exponent. To ensure that a fit was found, it was necessary to apply a bound to the plateau height search. In our analysis it was limited to a range from $0.2$ to $10$ times the mean of the first five data points of the window averaged PSD at low frequency. 


\subsection{Experiments}

Two experimental measurements with different period and sampling rate were performed at each temperature, resulting in two PSDs covering different frequency windows. These were combined to form PSD curves like those shown in Fig.~\ref{fig:trace}D in the main text, and the gaps in the data at $\nu\approx10^{3}$~Hz show where the datasets were joined. The PSDs were window averaged over all frequencies to reduce the noise. 

In the spin ice regime (temperatures between approximately $750$~mK and $1500$~mK) the Cole-Cole form gives a better fit to the experimental data (top panel of Fig.~\ref{appfig:intermediateTT}), with the Davidson-Cole fits generally overshooting the plateau at low frequency. 
At higher temperature (bottom panel of Fig.~\ref{appfig:highTT}), the shape of the curve evolves slowly, with the knee softening somewhat. As a result, the optimal fitting form becomes less clear, with both forms giving similarly good fits to the data. Notice that the dynamical range accessible in noise power decreases with the strength of the overall signal, and the uncertainty in the fitting parameters grows (see discussion in Sec.~\ref{sec:numdet_undersampling} below). 

One should notice that even though both Cole-Cole and Davidson-Cole forms give good fits, their extracted exponents differ by $\sim 15\%$ and the curves will separate if extended to larger frequencies. This difference provides an idea of the large systematic error bar on the measured exponent. 
Overall, the Cole-Cole form produces better fits across the broadest temperature range and therefore we chose not to present Davidson-Cole fits in the main text. 

\begin{center}\
\begin{figure}\
\includegraphics[width=0.95\columnwidth]{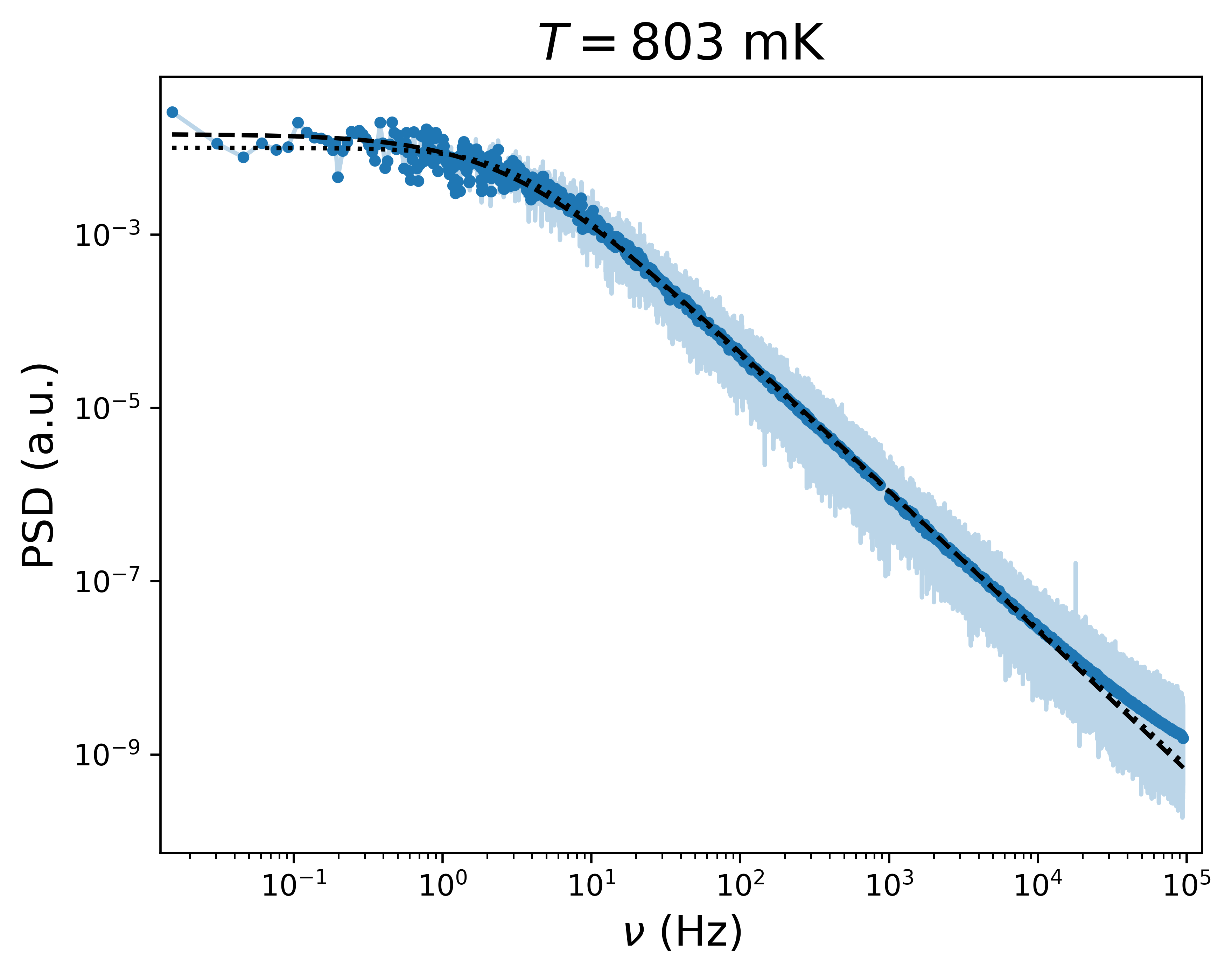}%
\\ 
\includegraphics[width=0.95\columnwidth]{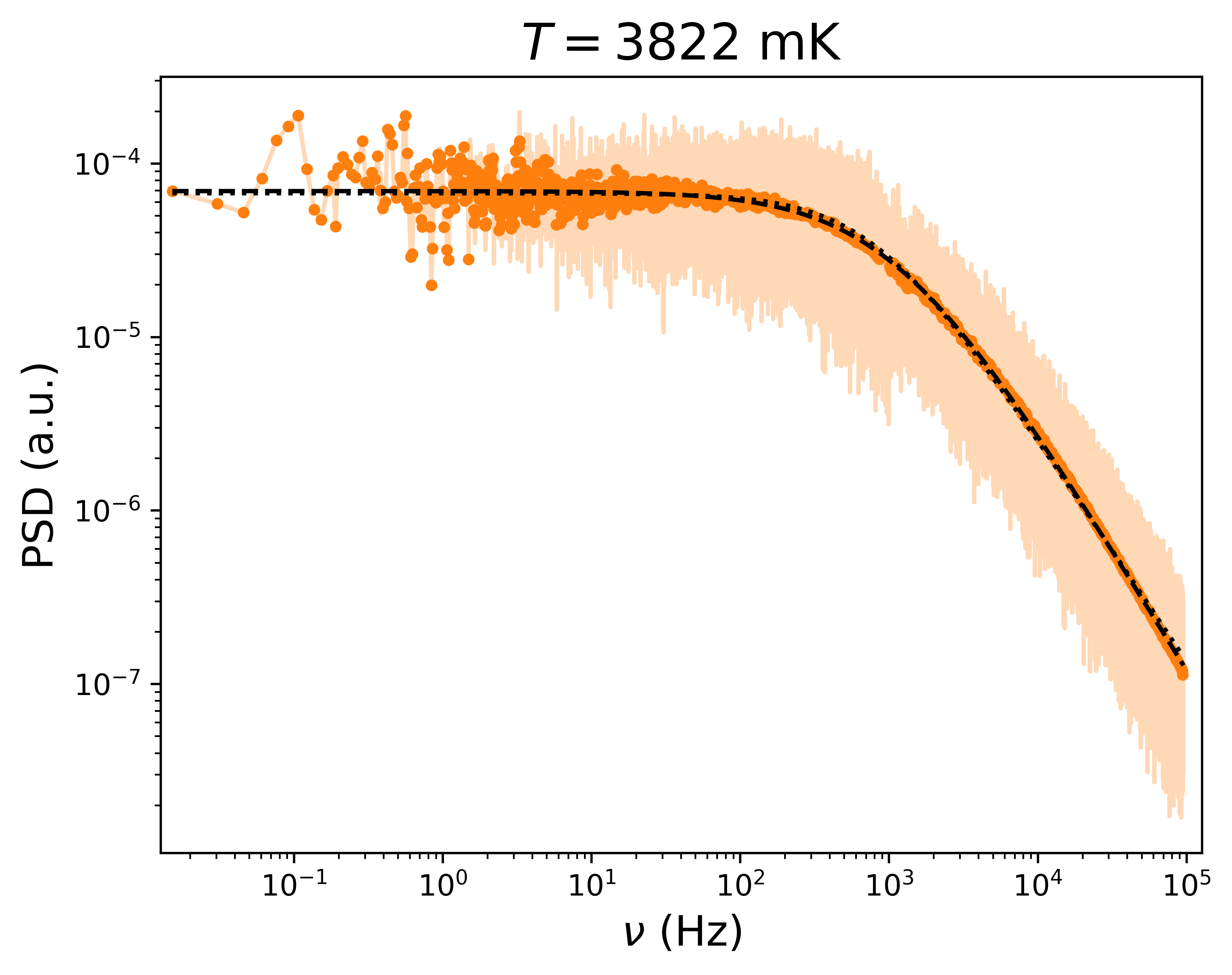}
\caption{\label{appfig:lowTT}\label{appfig:intermediateTT}\label{appfig:highTT}%
The raw (transparent lines) and window averaged (opaque points) PSD signal at two example temperatures.
Top panel: In the range between $750$~mK and $1.5$~K, a Cole-Cole (dotted black line) fits the data better than a Davidson-Cole (dashed black line) form. 
Bottom panel: At higher temperatures ($\gtrsim1.5$~K), the knee between plateau and scaling behaviour broadens. The high-frequency scaling regime is less clearly established and fitting for the exponent becomes more uncertain. Cole-Cole (dotted black) and Davidson-Cole (dashed black) fits give almost indiscernible curves.}\
\end{figure}\
\end{center}\


\subsection{Simulations}

\begin{figure}
    \centering
    \includegraphics[width=\columnwidth]{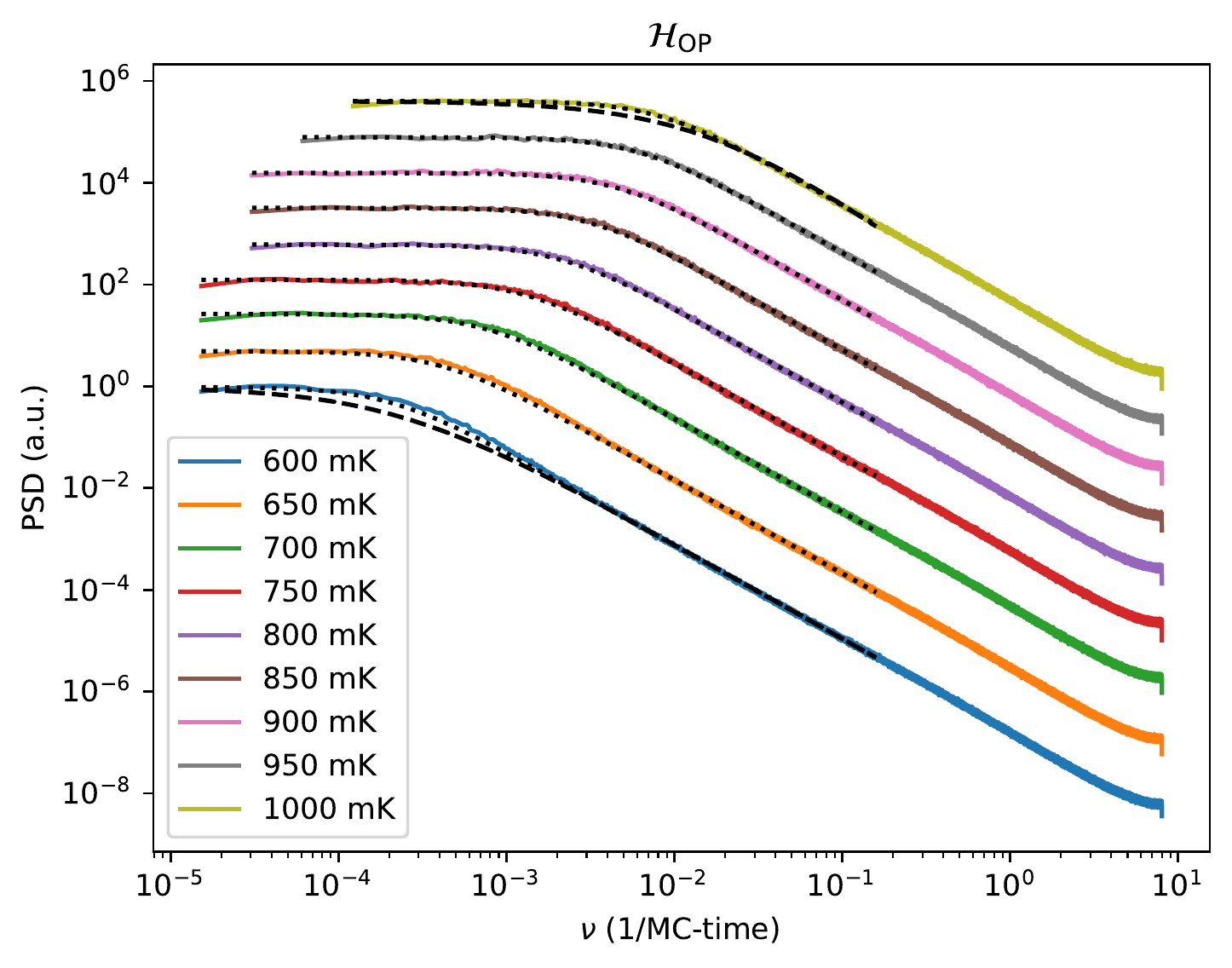}
    \caption{Monte Carlo data and fits for the ``optimal parameters'' (${\cal H}_{\mathrm{OP}}$), with Cole-Cole fits (dotted black lines). Davidson-Cole fits are shown for comparison only at the lowest and highest temperatures (dashed black lines). The Cole-Cole form is more consistent with the data, and the fitting parameters for it are shown in Fig.~\ref{fig:modelcomparison}B and \ref{fig:modelcomparison}C in the main text. The curves have been shifted vertically for clarity.}
    \label{fig:OPnumerics}
\end{figure}

For the simulation data, the Cole-Cole form was by far more consistent with the PSDs at all temperatures (see e.g., the example fits in Fig.~\ref{fig:OPnumerics}), and therefore we chose to discard the Davidson-Cole form altogether. 
The fitting was performed using the same procedure as for the experimental results. The PSDs from simulations were fitted over a frequency window $\nu_1< \nu < \nu_2$. In all cases $\nu_1$ was the smallest non-zero frequency available for the relevant set of data. As long as $\nu_1$ is chosen to lie firmly in the plateau regime, its value has no significant effect on the extracted parameters.
The choice of $\nu_2$ was rather more subtle. 
For $\nu \gg 1/(2\pi\tau_0)$, where $\tau_0=1$ MC step is the fastest microscopic timescale, the PSD always displays a decay with $\alpha=2$; this is because at the single spin flip level, MC implements a Poissonian process with a single characteristic time scale, leading to a corresponding short-time exponential decay of the autocorrelation function. 
This puts a limit on the largest sensible value for $\nu_2$. On the other hand, the fits were found to be unstable if one chose $\nu_2$ too close to the knee to the low frequency plateau. To enable a fair comparison between models and temperatures, we therefore chose to perform all our Cole-Cole fits up to $\nu_2=1/(2\pi)$ (MC step)$^{-1}$. This ensures that the fits cover as much of the power-law regime of the PSD as possible, avoiding any fits that terminate close to the knee, whilst not being dominated by the regime of (trivial) $\nu^{-2}$ decay at $\nu \gtrsim 1$ (MC step)$^{-1}$. For temperatures greater than $\sim 1.1$~K the knee shifts to frequencies close to $\nu_2=1/(2\pi)$ (MC step)$^{-1}$, and a stable fit is no longer possible. 

\begin{figure}
\centering
\includegraphics[width=\columnwidth]{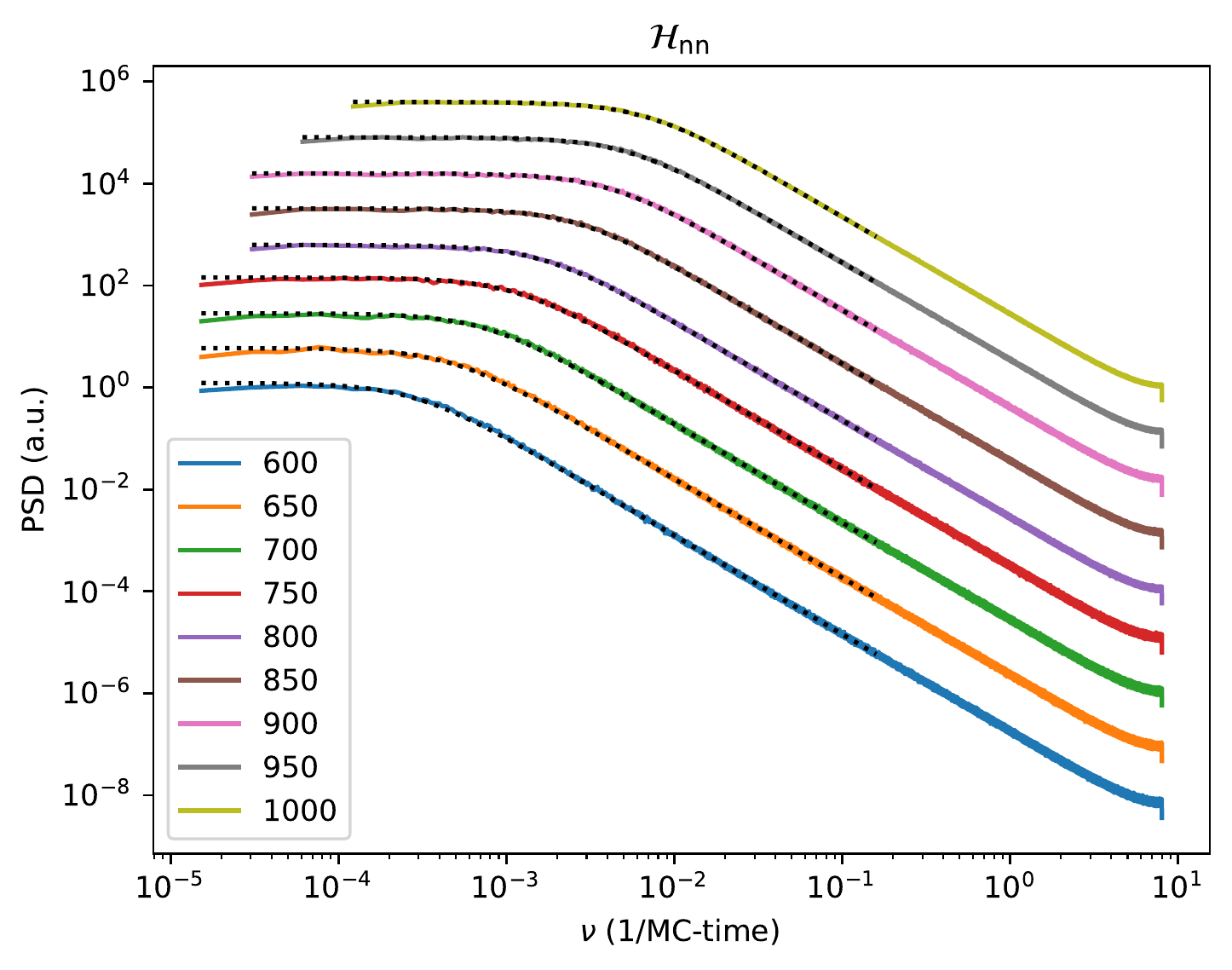}
\caption{Monte Carlo data and fits for the nearest neighbour model (${\cal H}_{\mathrm{nn}}$). Cole-Cole fits are denoted by dotted lines, with fitting parameters shown in Fig.~\ref{fig:modelcomparison}B and \ref{fig:modelcomparison}C in the main text. The curves have been shifted vertically for clarity.}
\label{fig:NNnumerics}
\end{figure}

\begin{figure}
\centering
\includegraphics[width=\columnwidth]{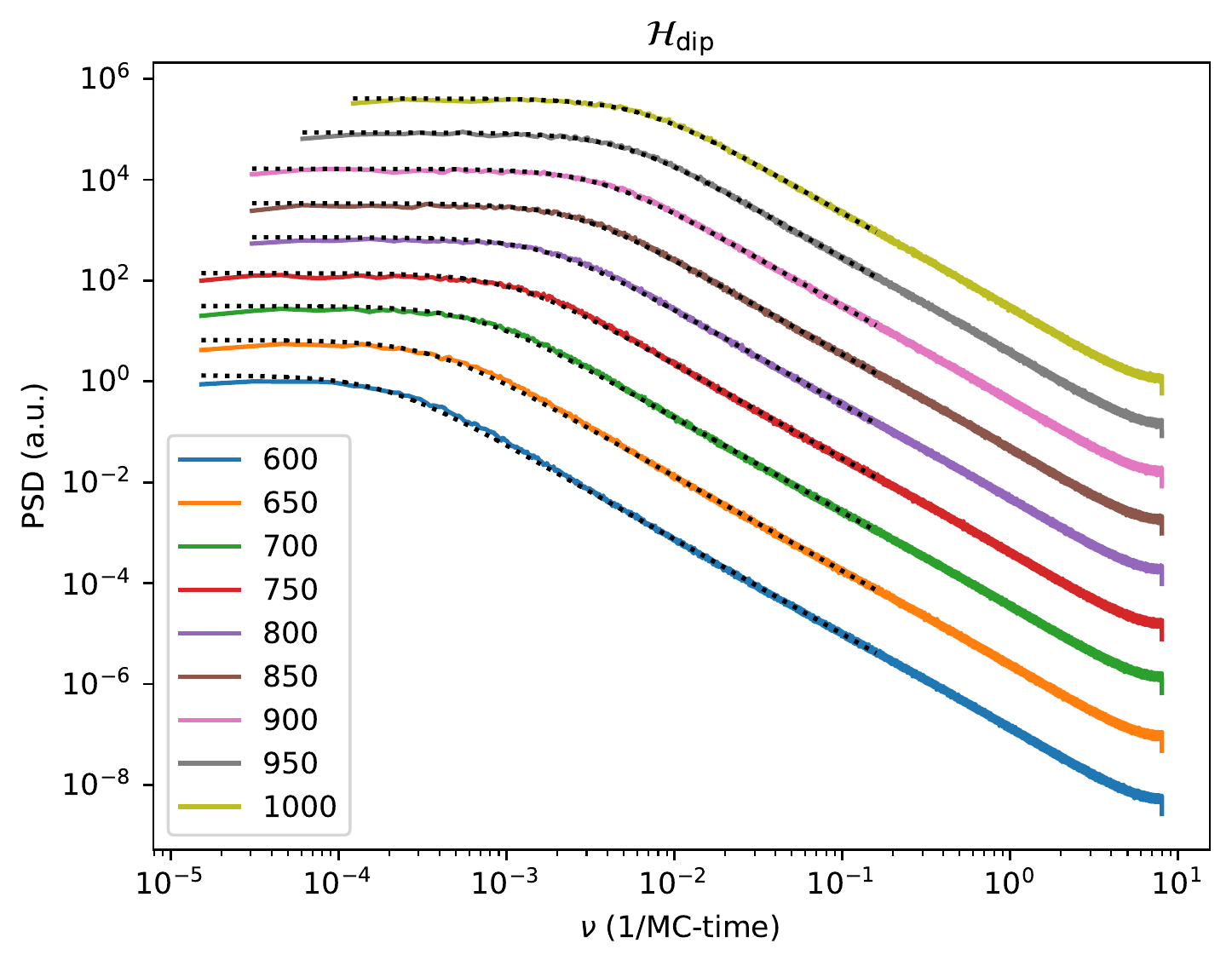}
\caption{Monte Carlo data and fits for the dipolar spin ice model (${\cal H}_{\mathrm{dip}}$). Cole-Cole fits are denoted by dotted lines, with fitting parameters shown in Fig.~\ref{fig:modelcomparison}B and \ref{fig:modelcomparison}C in the main text. The curves have been shifted vertically for clarity.}
\label{fig:DIPnumerics}
\end{figure}

\begin{figure}
\centering
\includegraphics[width=\columnwidth]{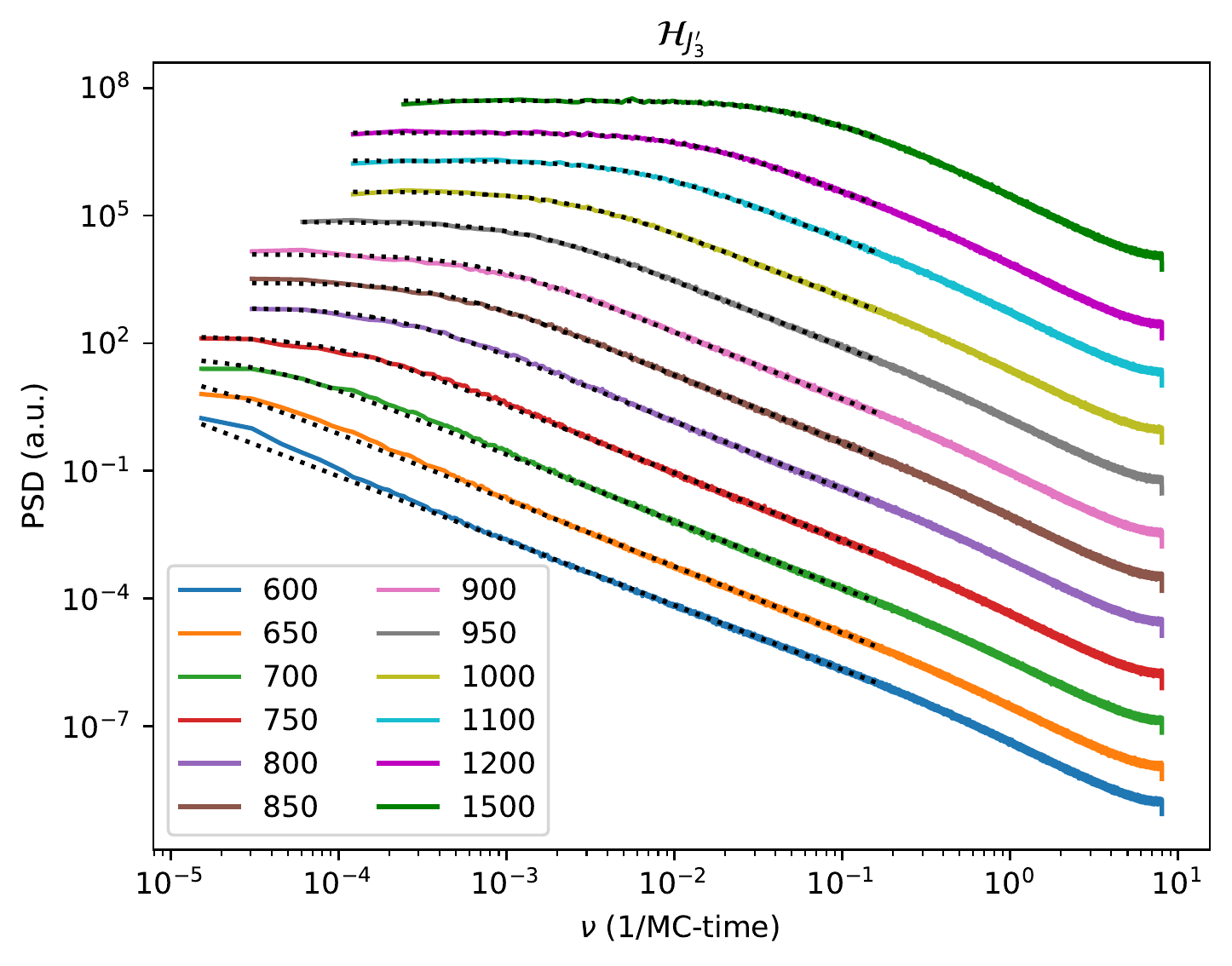}
\caption{Monte Carlo data and fits for the dipolar hamiltonian with $J_3'=0.4$~K (${\cal H}_{\mathrm{J_3^\prime}}$). Cole-Cole fits are denoted by dotted lines, with fitting parameters shown in Fig.~\ref{fig:modelcomparison}B and \ref{fig:modelcomparison}C in the main text. The curves have been shifted vertically for clarity.}
\label{fig:J3primeCCfits}
\end{figure}

The fits to the simulated PSDs of the Hamiltonians considered in this work are shown in Fig.~\ref{fig:NNnumerics},~\ref{fig:DIPnumerics},~\ref{fig:OPnumerics} and~\ref{fig:J3primeCCfits}. 
The results of the fits are shown in Fig.~\ref{fig:modelcomparison}B and \ref{fig:modelcomparison}C in the main text. 
%
%

%
%


\section{Drift of the anomalous exponent with $J_3'$}
\label{sec:numdet_drift}

\begin{figure}
\centering
\includegraphics[width=\columnwidth]{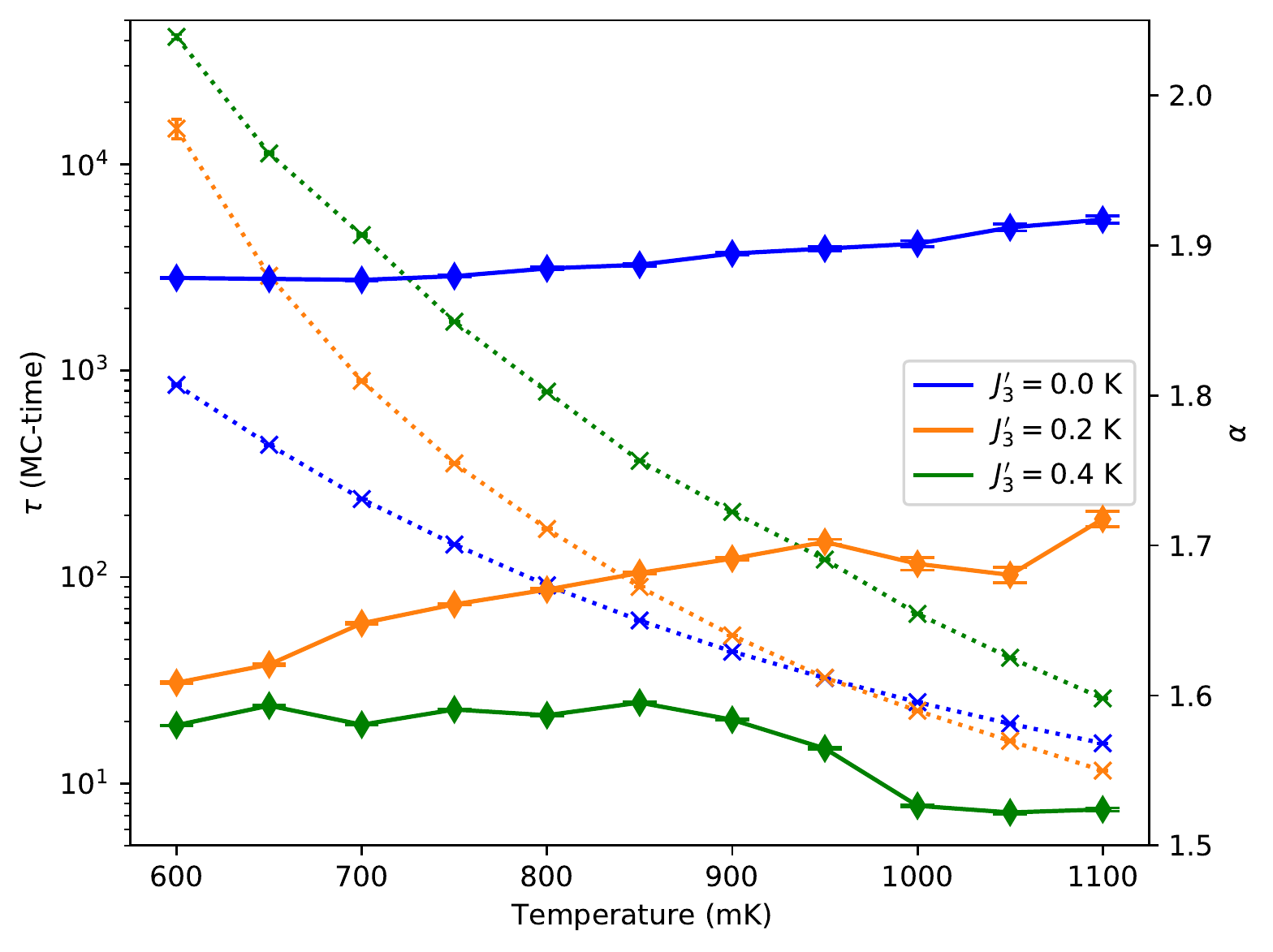}
\caption{\label{fig:j4drift}%
Characteristic relaxation time scale $\tau$ (dotted lines, left-hand-side  vertical  axis)  and  anomalous  exponent $\alpha$ (solid lines, right-hand-side vertical axis) for the model Hamiltonian with dipolar interactions and third-neighbour exchange across the hexagons with three different strengths of $J_3'$. Parameters  extracted from Cole-Cole fits to Monte Carlo data. The values $J_3'=0.0$~K and $J_3'=0.4$~K correspond to the Hamiltonians we have called ${\cal H}_{\rm{dip}}$ and ${\cal H}_{J_3'}$, respectively.}
\end{figure}

As the strength $J_3'$ of the third-neighbour exchange terms across the hexagonal loops is increased, the PSD curves go from being only weakly anomalous ($\alpha \simeq 2$) for $J_3'=0.0$~K (corresponding to ${\cal H}_{\rm{dip}}$, see Fig.~\ref{fig:DIPnumerics}) to being approximately as anomalous as the experimental results for $J_3'=0.4$~K (corresponding to ${\cal H}_{J_3'}$, see Fig.~\ref{fig:J3primeCCfits}) down to $T=600$~mK. The relaxation timescale also grows more rapidly as the temperature is decreased for large $J_3'$. For an intermediate strength $J_3'=0.2$~K at $T=1.1$~K the exponent $\alpha$ falls approximately halfway between the exponents for $J_3'=0.0$~K and $J_3'=0.4$~K, see Fig.~\ref{fig:j4drift}. 
However, at $T=0.6$~K the PSDs are almost as anomalous for $J_3'=0.2$~K as for $J_3'=0.4$~K. 

The relaxation timescale $\tau$ for $J_3'=0.2$~K  is comparable to (if not smaller than) for $J_3'=0.0$~K at high temperature, as shown in Fig.~\ref{fig:j4drift}. This is most likely because the antiferromagnetic $J_3'$ exchange counteracts the ferromagnetic dipolar interactions across the hexagonal loops, effectively reducing the average energy cost of flipping a spin at high temperature. As the temperature is reduced the non-zero third-neighbour exchange cause further ordering in the system, and generates a complex energy landscape for the magnetic monopoles to move through (see discussion in the main text).  This leads to reduced monopole mobility and a significantly larger relaxation time for nonzero $J_3'$ at low temperature.


\section{ \label{sec:numdet_undersampling}
Drift of the anomalous exponent from undersampling}

\begin{figure*}\
\includegraphics[width=\columnwidth]{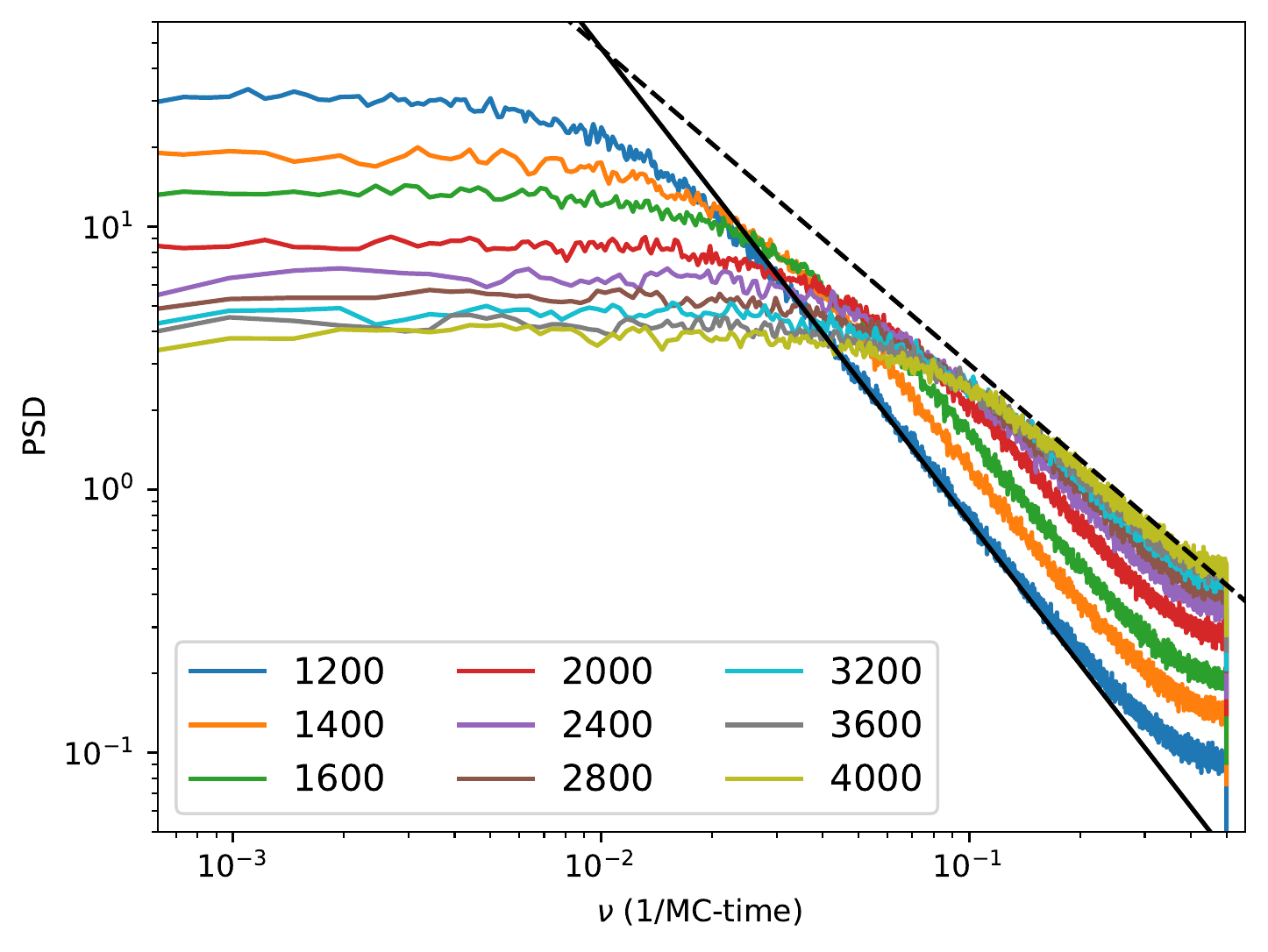}%
\includegraphics[width=\columnwidth]{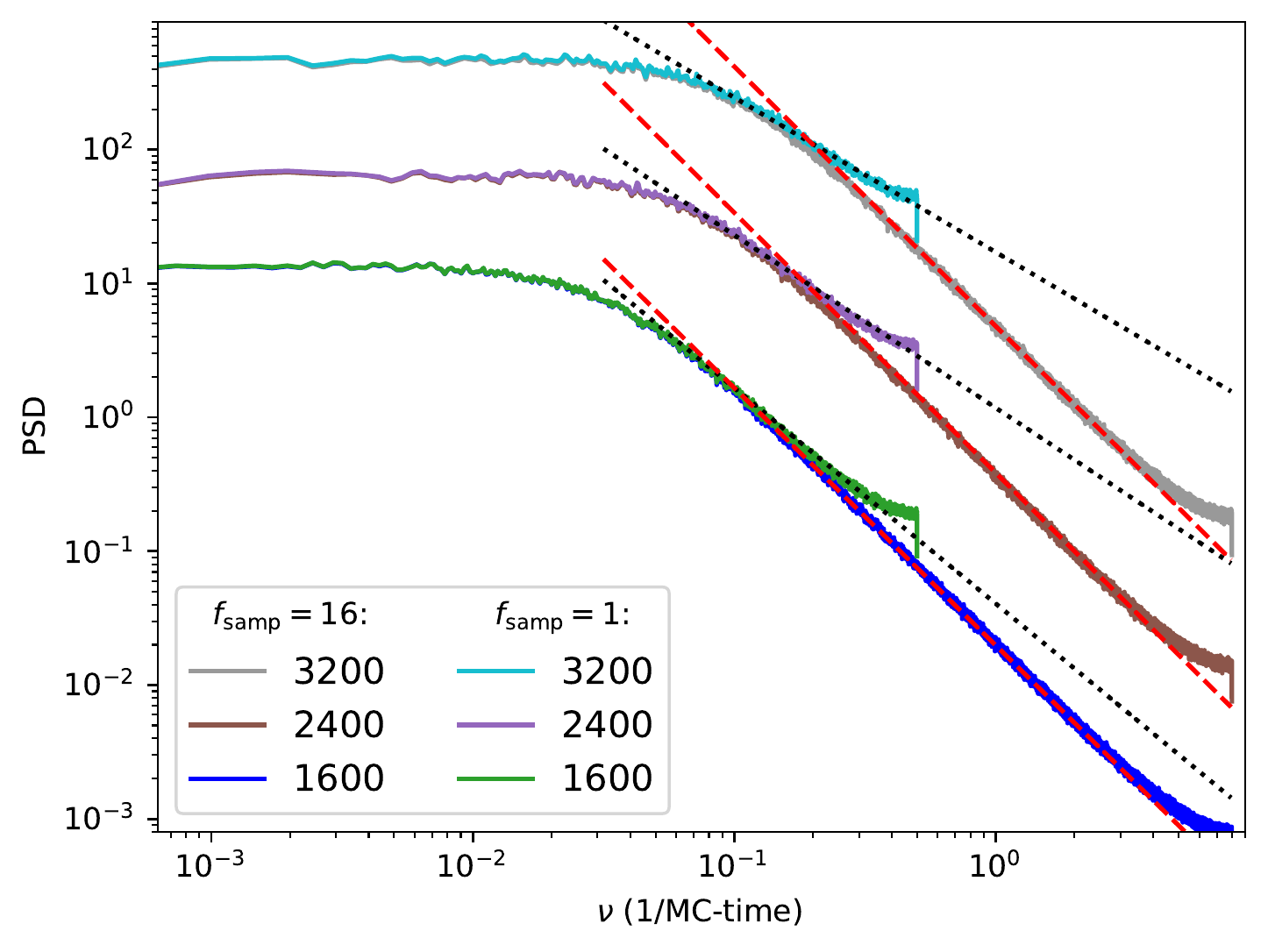}%
\caption{\label{fig:NNSampRate}%
This figure shows PSD curves from simulations of $\cal{H}_{\mathrm{dip}}$, and demonstrates the difficulty of deciding on a proper fitting procedure for Monte Carlo simulations at high temperature. The crossover from the low-frequency plateau to the high-frequency power law behaviour occurs close to the cutoff frequency $\nu \sim 1$ inverse MC steps. If we sample MC time in fractions of the MC step unit, we can access higher frequencies but this necessarily results in the $\nu^{-2}$ scaling discussed in Sec.~\ref{sec:numdet_fitting}. If instead we sample MC time in MC step units, any Cole-Cole or similar fits are deeply affected by the spurious high-frequency upturn in the PSD curves. This conundrum is illustrated in the figure with the aid of straight lines as guides to the eye. 
The legends indicate temperatures in mK.
Left: PSDs from data sampled once every Monte Carlo step for a selection of temperatures. The dashed and solid black lines are guides to the eye and correspond to $\nu^{1.2}$ and $\nu^{1.8}$, respectively.
Right: PSDs from simulations sampled either $16$ or $1$ time per Monte Carlo step at three temperatures. The dotted black and dashed red lines indicate the approximate slopes (extracted by eye) from the low and high sampling frequency curves. The exponents of the dotted black (dashed red) lines are, from top to bottom, $1.15$ ($1.94$), $1.29$ ($1.94$), and $1.61$ ($1.92$).}\
\end{figure*}

The upper frequency limit of the PSD is the sampling frequency. At the high frequency tail of the PSD calculated using Welch's method there is always a deviation from power-law behaviour due to aliasing effects. If one only samples the magnetisation once per MC step and then computes the PSD, these aliasing effects occur at $\nu\gtrsim 1$. Such sampling of the magnetisation data can hence lead to corruption of the high frequency data. At high temperatures the effect of undersampling can overlap with the knee at the end of the low frequency plateau, and result in a smaller exponent $\alpha$ if one attempts to perform a Cole-Cole fit to the data. This is demonstrated in Fig.~\ref{fig:NNSampRate}. 

%
%

%
%
\
%
%

\end{document}